\documentclass{article}

\usepackage[preprint]{neurips_2026}

\usepackage[utf8]{inputenc}
\usepackage[T1]{fontenc}
\usepackage{microtype}
\usepackage{graphicx}
\usepackage{url}
\usepackage{booktabs}
\usepackage{multirow}
\usepackage{nicefrac}
\usepackage{amsfonts}
\usepackage{amsmath}
\usepackage{amssymb}
\usepackage{xcolor}
\usepackage[table]{xcolor}
\usepackage{enumitem}
\usepackage{caption}
\usepackage{subcaption}
\usepackage{xspace}
\usepackage{wrapfig}
\usepackage{algorithm}
\usepackage{algpseudocode}

\usepackage[
  colorlinks=true,
  linkcolor={black!70!blue},
  citecolor={black!60!blue},
  urlcolor={black!50!blue},
  breaklinks=true
]{hyperref}
\usepackage{cleveref}

\crefname{section}{\S}{\S\S}
\Crefname{section}{Section}{Sections}
\crefname{subsection}{\S}{\S\S}
\Crefname{subsection}{Section}{Sections}
\crefname{table}{Tab.}{Tabs.}
\Crefname{table}{Table}{Tables}
\crefname{figure}{Fig.}{Figs.}
\Crefname{figure}{Figure}{Figures}
\crefname{appendix}{App.}{App.}
\Crefname{appendix}{Appendix}{Appendices}

\setlist[itemize]{topsep=2pt,itemsep=2pt,parsep=0pt,leftmargin=3.5mm}
\setlist[enumerate]{topsep=2pt,itemsep=2pt,parsep=0pt,leftmargin=3.5mm}

\newcommand{\our}{\textsc{TriProRep}\xspace}
\newcommand{\bench}{\textsc{RepSP}\xspace}
\usepackage{xcolor}
\usepackage[table]{xcolor}

\definecolor{deepblue}{HTML}{0A21BB}
\definecolor{lightblue}{HTML}{F2F3FF}

\title{Atom-level Protein Representation Learning \\
Improves Protein Structure Prediction}

\author{%
\textbf{
Taewon Kim$^{1,\ast}$ \quad
Hyosoon Jang$^{1,\ast}$ \quad
Hyunjin Seo$^{1}$ \quad
Seonghwan Seo$^{2}$ \quad
Hyeongwoo Kim$^{2}$
} \\[0.45em]
\textbf{
Wonho Zhung$^{2}$ \quad
Mingyeong Shin$^{2}$ \quad
Wooyoun Kim$^{2,3,4,5}$ \quad
Sungsoo Ahn$^{1,\dagger}$
} \\[0.65em]
KAIST \\
\texttt{\{maxkim139, hyosoon.jang, sungsoo.ahn\}@kaist.ac.kr}
}
\begin{document}

\maketitle
\begin{abstract}
Recent advances in generative modeling show that pretrained representations can improve generation as conditioning features or alignment targets. Motivated by this, we study protein representations for predicting structures beyond conventional function annotation. We propose \our{}, a structure-aware pretraining method that jointly models three aligned residue-level views: amino-acid identity, backbone geometry, and local full-atom geometry, discretely encoded via VQ-VAE tokenizers. By pretraining to recover original tokens from generator-corrupted views, \our{} learns to distinguish plausible but incorrect cross-view augmentations from the original protein. We further introduce \bench{}, a benchmark for evaluating protein representations in structure-predictive settings. \bench{} tests three uses of representations: homodimer co-folding from apo-chain representations, residue-level prediction of homodimer-derived interaction properties, and representation-aligned monomer structure prediction. Across these tasks, \our{} improves over sequence-only and prior structure-aware representation models, while maintaining competitive performance on conventional benchmarks.

\end{abstract}

\begingroup
\makeatletter
\renewcommand\@makefntext[1]{\noindent #1}
\makeatother
\renewcommand{\thefootnote}{}
\footnotetext{%
$^\ast$ Equal contribution. Author order was determined randomly. $^\dagger$ Corresponding author. $^1$ Graduate School of AI, KAIST; $^2$ Department of Chemistry, KAIST; $^3$ HITS Inc.; $^4$ Department of AX, KAIST; $^5$ InnoCORE AI-CRED Institute, KAIST.

\noindent
Code is available at \url{https://holymollyhao.github.io/TriProRep/}.%
}
\addtocounter{footnote}{-1}
\endgroup


\section{Introduction}
\label{sec:intro}
Structure-aware protein representation learning aims to produce residue-level features that can be reused by downstream models for three-dimensional reasoning such as complex prediction, interface modeling, and structure prediction. Yet this goal is only indirectly tested by many common protein representation benchmarks. Enzyme commission (EC) and gene ontology (GO) prediction~\citep{ec_and_go} measure broad biological utility, but they do not ask whether a representation exposes the geometric information needed to predict interfaces, assemble complexes, or supervise structure-prediction models. As a result, sequence-only representation models can remain competitive with explicitly structure-aware models on these benchmarks, which we also observe in \Cref{tab:ec-go-probing}, without resolving whether structural supervision has improved the representation for its intended use.

This mismatch is increasingly important as protein representation learning moves beyond sequence-only pretraining. SaProt~\citep{su2024saprot} and ProstT5~\citep{heinzinger2024bilingual} augment amino-acid sequences with Foldseek 3Di tokens~\citep{vankempen2024fast}. ESM3 represents proteins with coupled sequence, structure, and function tracks~\citep{hayes2025simulating}; and related methods condition sequence modeling on backbone structure or align sequence and structure representations~\citep{yang2023masked,wang2025s}. These approaches provide increasingly rich structural supervision, but their evaluation often remains centered on tasks that only weakly isolate residue-level geometry. A useful structure-aware representation should not merely improve functional annotation: it should provide features that downstream structure models can directly exploit.

This perspective mirrors a broader trend in representation learning. In computer vision, pretrained representations are increasingly used not only as task features, but also to improve structured-output models through conditioning~\citep{li2024return,ye2023ip-adapter,sereyjol2026test} and representation alignment-based distillation~\citep{yu2025representation,leng2025repa}. We ask the analogous question for proteins: 
\begin{center}
\noindent\fcolorbox{deepblue!30}{deepblue!3}{
\parbox{0.92\linewidth}{
\centering
\textcolor{deepblue!85!black}{\textbf{Question.}}
Can pretrained protein representations serve as useful geometric signals \\ for structure-predictive modeling?
}
}
\end{center}

\begin{figure*}[t]
\centering
\includegraphics[width=0.95\linewidth]{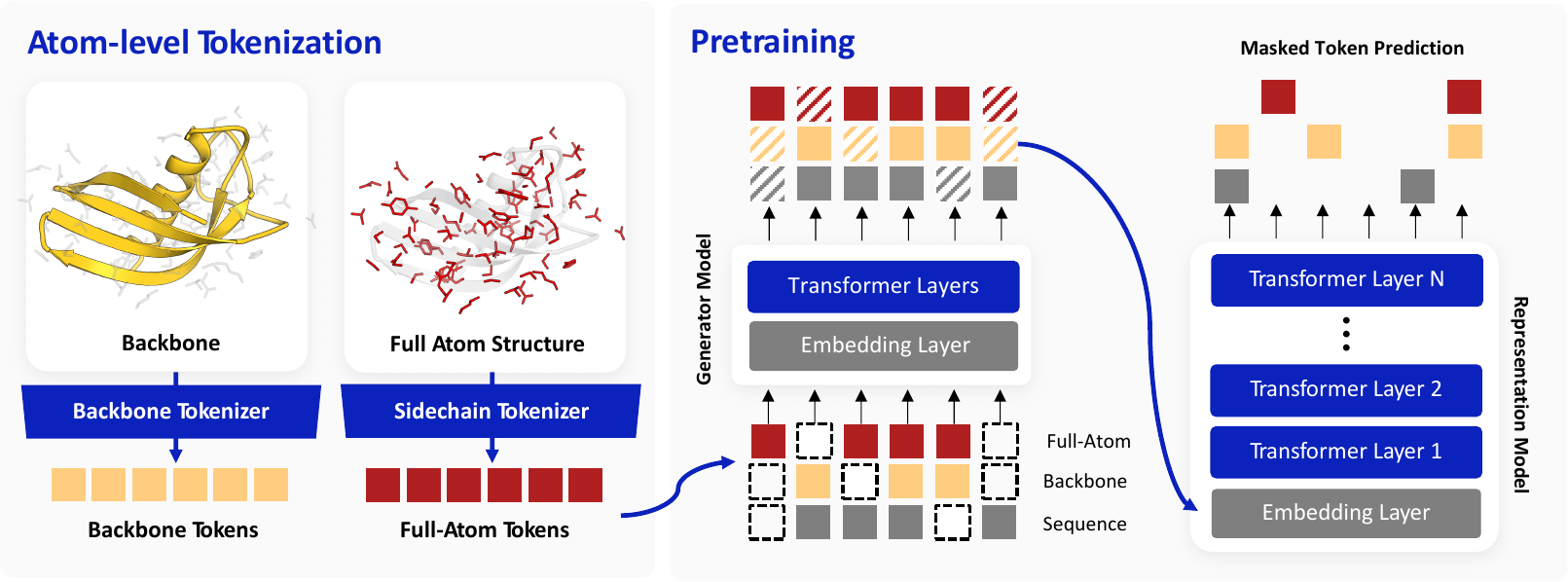}
\caption{\textbf{\our.} (a)~Three-view tokenization. A protein is independently tokenized into amino-acid, backbone, and full-atom token sequences at the residue level. (b)~ELECTRA-style discriminative pretraining. A small generator corrupts each of the three sequences, and a large discriminator predicts the original token at every position.}
\label{fig:method}
\vspace{-0.1in}
\end{figure*}

We study this question from both a methodological and evaluation perspective. Methodologically, we propose \our{}, a structure-aware protein representation model that jointly encodes three aligned residue-level views: amino-acid identity, backbone geometry, and full-atom residue geometry. The two geometric views are discretized by VQ-VAE tokenizers~\citep{van2017neural} into backbone-geometry and full-atom-geometry tokens. Unlike prior tokenizations that mainly capture backbone-level geometry~\citep{vankempen2024fast,hayes2025simulating,yuan2025protein}, our full-atom tokens preserve local atomic information, including side-chain arrangements that are important for residue environments and interfaces. To train these coupled views, we use corrective pretraining with generator-corrupted token sequences~\citep{clark2020electra}: a lightweight generator replaces masked tokens in each view with plausible alternatives, and the representation model recovers the original tokens.

For evaluation, we introduce \bench{}, the first benchmark specifically designed for structure-predictive protein representation learning. Built on homodimer complexes from the recent AFDB multimer release~\citep{han2026alphafold}, \bench{} provides a uniformly processed million-scale setting for controlled evaluation of binding-relevant geometry from apo-chain representations, while acknowledging that homodimers are simpler than general heteromeric complexes. \bench{} contains three complementary tasks: \emph{homodimer co-folding} from frozen apo-chain representations, \emph{homodimer-derived residue-level prediction} of binding sites, solvent-accessibility changes, interface regions, and interaction types, and \emph{representation-aligned structure prediction}, where pretrained representations serve as dense alignment targets for structure prediction learning~\citep{yu2025representation}.

We pretrain and evaluate \our{} at four scales, 35M, 150M, 650M, and 2.8B. On \bench{}, \our{} improves over sequence-only and prior structure-aware representation models across homodimer co-folding, homodimer-derived residue-level prediction, and representation-aligned structure prediction. On the EC/GO benchmarks, \our{} remains competitive with the strongest overall results, preserving broad biological representation quality. 

Our contributions are summarized as follows:
\begin{itemize}[topsep=-1.0pt,itemsep=0.5pt,leftmargin=3.5mm]
    \item We propose \our{}, a three-view structure-aware protein representation model trained with generator-corrupted token recovery.
    \item We introduce \bench{}, a benchmark that evaluates whether protein representations are useful for structure-predictive tasks.
    \item We show that \our{} improves homodimer co-folding, homodimer-derived residue-level prediction, and representation-aligned structure prediction, while remaining competitive on EC/GO.
\end{itemize}

\section{Related work}
\label{sec:related}

\textbf{Protein structure tokenization.}
A line of work uses VQ-VAE-based tokenizers \citep{van2017neural} to convert protein structures into discrete tokens for scalable structure-aware protein representation modeling. Foldseek defines 3Di tokens as a structural alphabet for residue-residue interaction geometry~\citep{vankempen2024fast}. ESM3 also uses tokens that discretize local backbone geometry into reconstructable tokens~\citep{hayes2025simulating}, and AminoAseed improves tokenizer quality through codebook reparameterization and Pareto-optimal codebook configuration~\citep{yuan2025protein}. However, these tokenizations mainly focus on backbone-level geometry and omit full-atom geometry within residues, such as side-chain packing, and rotameric states.

\textbf{Structure-aware protein representation learning.}
Recent protein representation models incorporate 3D structural information into protein language modeling. SaProt~\citep{su2024saprot} pairs amino-acid tokens with Foldseek’s 3Di tokens~\citep{vankempen2024fast}, and ProstT5~\citep{heinzinger2024bilingual} uses a T5 architecture to model bidirectional translation between sequence and 3Di tokens. ESM3~\citep{hayes2025simulating} introduces structure tokens alongside sequence and function tracks. In addition, MIF-ST~\citep{yang2023masked} performs masked inverse folding with backbone-structure conditioning, while S-PLM~\citep{wang2025s} aligns sequence and structure representations through contrastive learning. These methods show that structural information can improve protein representations. We study whether such representations transfer to structure-predictive tasks.

\textbf{Protein structure and complex prediction.}
Protein structure prediction has progressed from single-chain folding to protein-protein complex prediction. AlphaFold-Multimer~\citep{evans2021protein}, AlphaFold3~\citep{abramson2024accurate}, and Boltz~\citep{wohlwend2025boltz,passaro2025boltz2} predict protein complexes using MSA-derived representations, while SimpleFold~\citep{wang_simplefold_2025} revisits single-chain folding with flow matching and a general-purpose Transformer architecture using protein sequence representation. Our work uses structure prediction as an evaluation setting, rather than proposing a new structure prediction model. We evaluate whether structure-aware representations extracted from a single apo protein support flexible-docking prediction of the protein complex and serve as distillation targets for training single-protein folding models.

\textbf{Representation-based structure modeling.} In computer vision, structured-output models often use pretrained representations as conditioning signals~\citep{li2024return,ye2023ip-adapter,sereyjol2026test} or distillation sources~\citep{yu2025representation,leng2025repa}. Analogously, protein structure prediction models such as ESMFold~\citep{lin2023evolutionary} and SimpleFold~\citep{wang_simplefold_2025} use sequence-only protein representations, particularly ESM2 representations~\citep{lin2023evolutionary}. In this paper, we study structure-aware protein representations in both roles: as input conditioning signals for flexible protein-protein docking, and as dense distillation sources for monomer structure prediction.


\section{\our: A three-view protein structure representation}
\label{sec:method}

In this section, we introduce \our{}, a structure-aware protein representation model that jointly encodes three residue-level views of a protein: amino-acid identity, backbone geometry, and full-atom geometry. We first describe how these views are converted to discrete tokens (\Cref{sec:method_tokens}). We then describe a corrective pretraining objective over generator-corrupted token sequences (\Cref{sec:method_pretrain}). The full pipeline is illustrated in \Cref{fig:method}, with implementation details in \Cref{appx:method}.

\subsection{Three-view residue tokenization}
\label{sec:method_tokens}

Given a protein, we construct three aligned residue-level token sequences. For each residue, we define an amino-acid token, a backbone-geometry token, and a full-atom-geometry token. These tokens provide three complementary views of the same residue: sequence identity, local backbone geometry, and intra-residue full-atom geometry. The amino-acid token encodes residue identity, as in sequence-only protein language models such as ESM2~\citep{lin2023evolutionary}. We pair this sequence-identity view with two structural views, each defined at the residue level.

\textbf{Backbone-geometry tokens.}
The backbone-geometry token provides a residue-level structural view of the local protein backbone. We obtain these tokens using AminoAseed~\citep{yuan2025protein}, a VQ-VAE-based tokenizer that maps local backbone substructures around each residue into discrete codes. Compared with Foldseek 3Di tokens used by a prior work \citep{su2024saprot} that also performs residue backbone tokenization, AminoAseed improves codebook utilization and token diversity, providing a more expressive backbone tokenization. 

\textbf{Full-atom-geometry tokens.} We newly introduce the full-atom-geometry token to encode intra-residue full-atom geometry, complementing the backbone-geometry token with local atomic information within each residue (as shown in \Cref{fig:tokenizer_comparison} of \Cref{appx:fullatom_token}). The full-atom-geometry token focuses on heavy-atom arrangements in a backbone-defined local frame, including side-chain rotamer geometry. This complementary tokenization provides information that is largely inaccessible from backbone geometry alone, allowing the model to incorporate residue-level atomic details that are useful for downstream protein-geometry understanding. We obtain this token by training a residue-level VQ-VAE tokenizer that takes heavy-atom coordinates expressed in an SE(3)-invariant local frame defined by the $N$, $\mathrm{C}\alpha$, and $C$ atoms, together with dihedral-angle features, and assigns each residue a categorical code. We provide details in \Cref{appx:fullatom_token}.

\subsection{Corrective pretraining with generator-corrupted views}\label{sec:method_pretrain}

We pretrain \our{} using a corrective token-recovery objective over the three aligned token sequences. The objective is inspired by ELECTRA-style pretraining~\citep{clark2020electra}, but differs in that the representation model predicts the original token rather than a binary replacement label. In our three-view setting, generator-based replacement serves as token-level augmentation. Since each residue is represented by amino-acid, backbone-geometry, and full-atom-geometry tokens, independent replacement can produce inputs that are plausible within each view but mutually inconsistent across views. Recovering the original tokens encourages the representation model to model consistency among sequence identity, backbone geometry, and residue full-atom geometry.


At each step, we independently sample masked positions for each token type. We feed the partially masked amino-acid, backbone-geometry, and full-atom-geometry token sequences to a small generator. The generator is trained with masked-token cross-entropy to predict the original token distribution at each masked position. We then sample replacement tokens from the generator distributions and construct generator-corrupted token sequences.

The representation model takes the corrupted token sequences as input and predicts the original token at every position for all three token types, using separate token-recovery heads for the three views. We use the corrective objective~\citep{xu2020mc}, which recovers the original token rather than predicting a binary replaced-or-unchanged label. This provides dense supervision at every residue position and exposes the model to a large set of generator-produced cross-view corruptions. After pretraining, we discard both the generator and the token-recovery heads, and use the hidden states of the representation model as the transferable protein representation. We provide details in \Cref{appx:pretrain}.

\textbf{Datasets.} We pretrain \our{} on 83.7M predicted protein structures from the AlphaFold Protein Structure Database (AFDB)~\citep{varadi2022alphafold} and ESMAtlas~\citep{lin2023evolutionary}. AFDB provides AlphaFold2-predicted structures over proteome-scale sequences in UniProt~\citep{uniprot2023uniprot}, and ESMAtlas provides ESMFold predictions over metagenomic sequences in MGnify~\citep{mitchell2020mgnify} augmenting the AFDB structures. For AFDB, we use the representative protein from each UniRef30 cluster and retain structures with pLDDT at least 50, yielding 18.2M representative proteins. For ESMAtlas, we retain structures with pLDDT greater than 70, yielding 65.5M additional predicted structures.
\section{A benchmark for structure-predictive protein representation learning}
\label{sec:exp_setup}

\begin{figure}[!t]
\centering
\includegraphics[width=\linewidth]{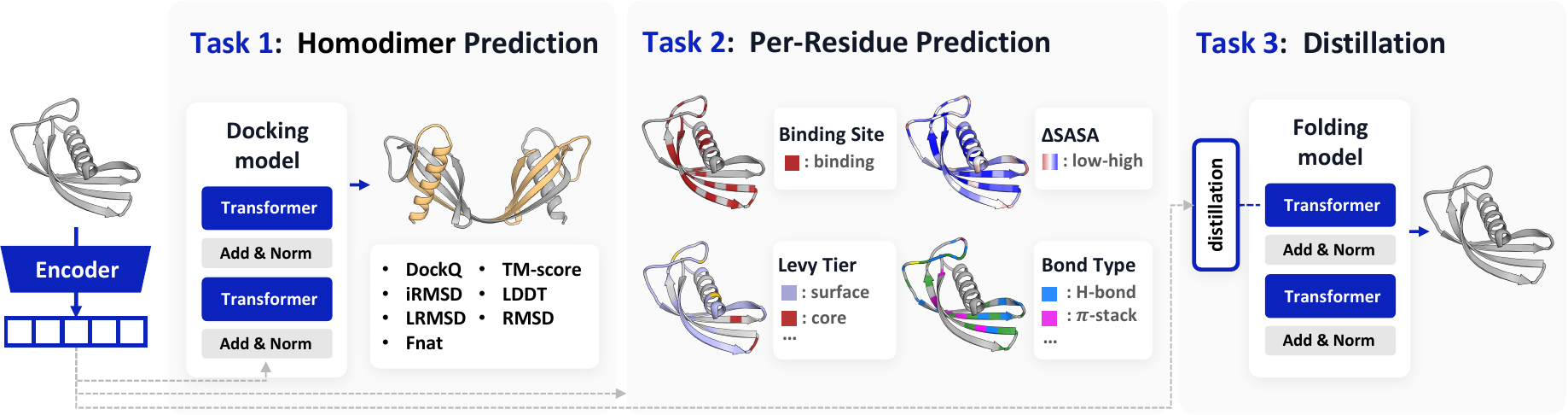}
\caption{\textbf{\bench.} We define three structure-generative tasks that use protein representations as input: \textbf{(task 1)} homodimer structure prediction, \textbf{(task 2)} per-residue homodimer binding-property prediction via MLP probing, and \textbf{(task 3)} distillation into a monomer structure prediction model.
}\label{fig:benchmark}
\end{figure}

We introduce \bench{}, which evaluates whether protein representations improve structure prediction across three tasks: (1) inferring homodimer complex structures, (2) predicting residue-level labels derived from homodimer structures, and (3) accelerating monomer protein structure prediction via distillation \citep{yu2025representation}. To the best of our knowledge, our work is the first to use structure-aware representations for flexible-docking prediction and distillation. The full pipeline is illustrated in \Cref{fig:benchmark}, with implementation details in \Cref{appx:benchmark}.

\textbf{Dataset curation and split.}
This benchmark uses $1.8$M homodimer complexes released in the March 2026 AlphaFold Protein Structure Database (AFDB)~\citep{han2026alphafold}. We focus on homodimers because they are currently the largest uniformly processed complex dataset in AFDB, providing a controlled setting for evaluation at scale: the binding partner is identical to the query chain, thereby removing variation along the partner-pairing axis while still requiring interface-relevant geometric information.\footnote{We also evaluate representations in heterodimer settings, as shown in \Cref{tab:binder-screening-binding-site} of \Cref{sec:abblation}.}


To be specific, we pair each holo homodimer complex with its corresponding monomer protein prediction in the AFDB~\citep{varadi2022alphafold,varadi2024alphafold,jumper2021highly}. We then split the dataset by sequence clusters at $50\%$ sequence identity. From the resulting cluster representatives, we select 400 validation and 1{,}000 test sequences, and use the remaining representatives for training. We provide details in \Cref{appx:benchmark_curation}.

\textbf{Homodimer structure prediction (\Cref{sec:exp_cofolding}).} We evaluate whether protein representations are predictive of the corresponding homodimer structure. Strong performance indicates that the protein representation contains structural information useful for protein complex structure inference. Specifically, a $100$M flexible-docking model takes monomer protein representations as input and learns to predict the homodimer structures. We modify SimpleFold~\citep{wang_simplefold_2025}, a Transformer-based single-protein folding model~\citep{lin2023evolutionary}, into a homodimer flexible-docking model, with details provided in \Cref{appx:homodimer_prediction}. We evaluate the predicted structures using interface-quality metrics, including DockQ, iRMSD, LRMSD, and Fnat, and overall quality metrics, including TM-score, LDDT, and RMSD.

\textbf{Per-residue homodimer binding property prediction (\Cref{sec:exp_homomer}).} We next evaluate whether monomer protein representations are predictive of residue-level binding properties of the corresponding homodimer complex. Strong probing performance indicates that the representation captures binding-relevant local signals, including binding residues, exposure changes, interface regions, and interaction types. Given residue-wise monomer representations, we use MLP probing with four per-residue targets: {binary binding site}, continuous change in solvent-accessible surface area between the monomer and homodimer structures ({$\Delta$SASA})~\citep{lee1971interpretation,shrake1973environment}, {Levy tier} with five classes, and multi-label {bond type} over five interaction classes~\citep{salentin2015plip,jubb2017arpeggio}. We describe details in \Cref{appx:benchmark_perresidue}.

\textbf{Distillation into monomer structure prediction (\Cref{sec:exp_folding}).} We evaluate whether protein representations provide useful distillation targets for monomer structure prediction. This experiment is motivated by recent work showing that high-quality representations can improve generative model training through distillation~\citep{yu2025representation}. Following this, we distill monomer protein representations into the SimpleFold-$100$M model~\citep{wang_simplefold_2025} by maximizing the cosine similarity between monomer protein representations and the model's hidden states for structure prediction. Implementation details are provided in \Cref{appx:benchmark_monomer}. We assess representation quality by the resulting improvement in structure prediction, using TM-score, GDT-TS, GDT-HA, LDDT, backbone-based LDDT$_{\text{bb}}$, and RMSD.

\section{Results}
\label{sec:experiments}

We pretrain \our at four model scales: 35M, 150M, 650M, and 2.8B. Then, we compare against sequence-only ESM2~\citep{lin2023evolutionary}, structure-aware representation models, including SaProt~\citep{su2024saprot}, S-PLM~\citep{wang2025s}, MIF-ST~\citep{yang2023masked}, ESM3~\citep{hayes2025simulating}, and ProstT5~\citep{heinzinger2024bilingual}. We report on \bench{}, a new benchmark that evaluates the structure-generative capacity of protein representations (\Cref{sec:exp_cofolding,sec:exp_homomer,sec:exp_folding}), and on conventional protein representation benchmarks (\Cref{sec:abblation}). 

\begin{table}[t]
\centering
\caption{\textbf{Homodimer flexible-docking performance on \bench{}.} Monomer protein representations predict homodimer structures. \textbf{Bold} denotes the best within each size class. Overall, flexible-docking with representations from \our{} outperforms the baselines across all parameter scales.}
\resizebox{.95\linewidth}{!}{
\setlength{\tabcolsep}{4pt}
\renewcommand{\arraystretch}{0.9}
\label{tab:cofolding-test}
\begin{tabular}{lc|cccc|ccc}
\toprule
& & \multicolumn{4}{c|}{\textbf{Interface quality}}
& \multicolumn{3}{c}{\textbf{Overall quality}} \\
\textbf{Models}
& \textbf{Params} & DockQ $\uparrow$ & iRMSD $\downarrow$ & LRMSD $\downarrow$ & Fnat $\uparrow$
& TM-score $\uparrow$ & LDDT $\uparrow$ & RMSD $\downarrow$ \\
\hline

\rowcolor{gray!12}\multicolumn{9}{l}{\textit{Small models}} \\
ESM2     & 35M  & 0.278          & 11.549         & 25.942         & 0.417          & 0.531          & 0.603          & 16.176 \\
SaProt   & 35M  & 0.296          & 10.380         & 25.308         & 0.414          & 0.567          & 0.712          & 14.660 \\
\rowcolor{lightblue}
\textcolor{deepblue}{\textbf{Ours}} & 35M  & \textbf{0.371} & \textbf{8.915} & \textbf{21.964}& \textbf{0.486} & \textbf{0.633} & \textbf{0.771} & \textbf{12.778} \\

\hline
\rowcolor{gray!12}\multicolumn{9}{l}{\textit{Medium models}} \\
ESM2     & 150M & 0.310          & 10.743         & 25.328         & 0.450          & 0.563          & 0.667          & 15.170 \\
\rowcolor{lightblue}
\textcolor{deepblue}{\textbf{Ours}} & 150M & \textbf{0.419} & \textbf{8.170} & \textbf{20.292}& \textbf{0.512} & \textbf{0.666} & \textbf{0.812} & \textbf{11.874} \\

\hline
\rowcolor{gray!12}\multicolumn{9}{l}{\textit{Large models}} \\
ESM2     & 650M  & 0.374          & 9.517          & 22.803         & 0.533         & 0.613          & 0.721          & 13.913 \\
SaProt   & 650M  & 0.443          & 7.449          & 18.878         & 0.548         & 0.690          & 0.807          & 11.196 \\
S-PLM    & 704M  & 0.366          & 9.350          & 22.678         & 0.507         & 0.619          & 0.708          & 13.591 \\
MIF-ST   & 643M  & 0.299          & 10.598         & 25.420         & 0.384         & 0.563          & 0.818          & 14.521 \\
\rowcolor{lightblue}
\textcolor{deepblue}{\textbf{Ours}} & 650M  & \textbf{0.477} & \textbf{6.883} & \textbf{18.243}& \textbf{0.588}& \textbf{0.705} & \textbf{0.829} & \textbf{10.554} \\

\hline
\rowcolor{gray!12}\multicolumn{9}{l}{\textit{Huge models}} \\
ESM2     & 3B    & 0.387          & 8.679          & 21.455         & 0.547          & 0.635          & 0.732          & 13.025 \\
ESM3     & 1.4B  & 0.448          & 7.060          & 18.554         & 0.579          & 0.695          & \textbf{0.835} & 10.719 \\
ProstT5  & 1.2B  & 0.409          & 7.790          & 20.403         & 0.541          & 0.661          & 0.782          & 11.843 \\
\rowcolor{lightblue}
\textcolor{deepblue}{\textbf{Ours}} & 2.8B    & \textbf{0.499} & \textbf{6.370} & \textbf{17.087}& \textbf{0.612} & \textbf{0.724} & 0.834          & \textbf{9.875}  \\
\bottomrule
\end{tabular}}
\end{table}
\begin{figure}[t]
\centering
\vspace{-7pt}
\includegraphics[width=.9\linewidth]{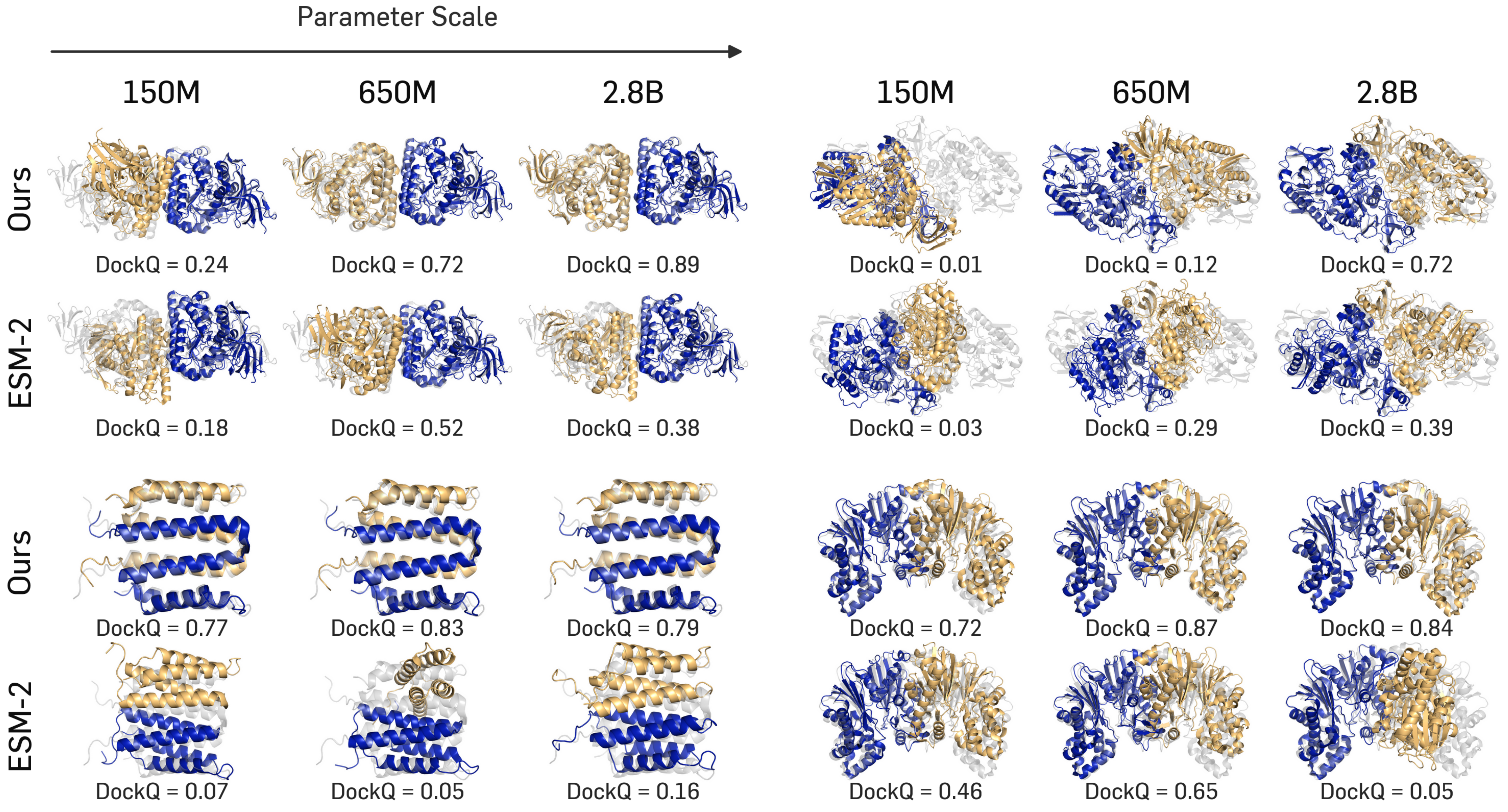}
\caption{\textbf{Scaling of flexible-docking.} Predicted homodimer structures (chain A blue, chain B gold) overlaid on ground truth (gray) across encoder sizes (150M, 650M, 3B) for {four} test records.}
\label{fig:cofolding-scale}
\vspace{-.1in}
\end{figure}

\subsection{Homodimer structure prediction}
\label{sec:exp_cofolding}
\Cref{tab:cofolding-test} reports homodimer flexible-docking performance on \bench{}, with qualitative examples shown in \Cref{fig:cofolding-scale}. Across scales from 35M to 2.8B parameters, \our{} consistently improves both interface quality and overall structural accuracy, indicating favorable scaling with model size. Notably, the 650M \our{} model already outperforms the huge baselines on all reported metrics, while the huge \our{} model achieves the strongest performance on nearly all metrics, with ESM3 only marginally higher in LDDT.

The gains are most pronounced on interface-level metrics, which depend not only on accurate monomer structures but also on their relative placement and contact geometry. The qualitative examples in \Cref{fig:cofolding-scale} show the same trend: larger \our{} models reliably recover the correct monomer orientation, whereas baseline predictions often misplace the interface. These results suggest that \our{} provides representations that better support complex-level geometric inference.

Our \bench{} benchmark also highlights the importance of structural information beyond parameter count. Structure-aware models, including SaProt-650M, ESM3-1.4B, and ProstT5-1.2B, outperform the larger sequence-only ESM2-3B on most metrics. This pattern suggests that homodimer flexible docking is a structure-sensitive task, making \bench{} a useful evaluation setting for structure-dependent representation quality.



\begin{table}[t]
\centering
\caption{\textbf{Per-residue homodimer binding properties probing on \bench.} MLP probes on monomer protein representations predict properties on the homodimer. \textbf{Bold} indicates the best within each size class. Results are averaged over three runs. Our method shows strongest performance.}
\label{tab:homomer-probing}
\resizebox{.95\linewidth}{!}{
\setlength{\tabcolsep}{4pt}
\renewcommand{\arraystretch}{0.95}
\begin{tabular}{lc|cc|cc|cc|cc}
\toprule
& & \multicolumn{2}{c|}{\textbf{Binding site}}
& \multicolumn{2}{c|}{\textbf{$\Delta$SASA}}
& \multicolumn{2}{c|}{\textbf{Levy tier}}
& \multicolumn{2}{c}{\textbf{Bond type}} \\
\textbf{Models} & \textbf{Params}
& AUPRC & AUROC 
& Pearson & Spearman 
& Macro F1 & Acc. 
& AUPRC & AUROC \\
\hline

\rowcolor{gray!12}\multicolumn{10}{l}{\textit{Small}} \\
ESM2    & 35M    & 0.6635 & 0.8862 & 0.5955 & 0.3445 & 0.5386 & 0.7706 & 0.2936 & 0.9338 \\
SaProt  & 35M    & 0.6291 & 0.8794 & 0.6218 & 0.3923 & 0.5248 & 0.8036 & 0.2663 & 0.9388 \\
\rowcolor{lightblue}
\textcolor{deepblue}{\textbf{Ours}} & 35M    & \textbf{0.7614} & \textbf{0.9243} & \textbf{0.7047} & \textbf{0.4167} & \textbf{0.6260} & \textbf{0.8342} & \textbf{0.3566} & \textbf{0.9518} \\

\hline
\rowcolor{gray!12}\multicolumn{10}{l}{\textit{Medium}} \\
ESM2    & 150M   & 0.7087 & 0.9062 & 0.6379 & 0.3703 & 0.5665 & 0.7969 & 0.3227 & 0.9425 \\
\rowcolor{lightblue}
\textcolor{deepblue}{\textbf{Ours}} & 150M   & \textbf{0.8161} & \textbf{0.9433} & \textbf{0.7801} & \textbf{0.4671} & \textbf{0.6843} & \textbf{0.8668} & \textbf{0.4204} & \textbf{0.9642} \\

\hline
\rowcolor{gray!12}\multicolumn{10}{l}{\textit{Large}} \\
ESM2    & 650M   & 0.7915 & 0.9369 & 0.7146 & 0.4113 & 0.6442 & 0.8281 & 0.3902 & 0.9566 \\
SaProt  & 650M   & 0.8273 & 0.9508 & 0.7881 & 0.4715 & 0.6946 & 0.8693 & 0.4305 & \textbf{0.9692} \\
S-PLM   & 704M   & 0.7789 & 0.9324 & 0.6967 & 0.4036 & 0.6215 & 0.8166 & 0.3602 & 0.9523 \\
MIF-ST  & 643M   & 0.5846 & 0.8671 & 0.6726 & 0.4338 & 0.5261 & 0.8435 & 0.2623 & 0.9430 \\
\rowcolor{lightblue}
\textcolor{deepblue}{\textbf{Ours}} & 650M   & \textbf{0.8524} & \textbf{0.9561} & \textbf{0.8088} & \textbf{0.4813} & \textbf{0.7290} & \textbf{0.8780} & \textbf{0.4425} & 0.9673 \\

\hline
\rowcolor{gray!12}\multicolumn{10}{l}{\textit{Huge}} \\
ESM2    & 3B     & 0.8247 & 0.9483 & 0.7446 & 0.4281 & 0.6851 & 0.8426 & 0.4106 & 0.9614 \\
ESM3    & 1.4B   & 0.8470 & 0.9558 & 0.8121 & \textbf{0.4878} & 0.7285 & \textbf{0.8898} & 0.4229 & 0.9673 \\
ProstT5 & 1.2B   & 0.8291 & 0.9480 & 0.7574 & 0.4449 & 0.6848 & 0.8538 & 0.4077 & 0.9594 \\
\rowcolor{lightblue}
\textcolor{deepblue}{\textbf{Ours}} & 2.8B   & \textbf{0.8650} & \textbf{0.9608} & \textbf{0.8152} & 0.4854 & \textbf{0.7408} & 0.8811 & \textbf{0.4516} & \textbf{0.9691} \\

\bottomrule
\end{tabular}}
\vspace{-.15in}
\end{table}


\subsection{Per-residue homodimer binding property prediction}
\label{sec:exp_homomer}
\Cref{tab:homomer-probing} reports per-residue probing results on \bench{}. Across all model scales, \our{} improves over the baselines on most tasks and metrics, with the strongest overall performance attained at the huge scale. The consistency of these gains indicates that the benefits of the proposed representation are not confined to a particular model size, but scale reliably across the model family.

These probing results offer a residue-level explanation for the co-folding improvements in \Cref{sec:exp_cofolding}. Since the probes are trained on frozen monomer representations, performance gains reflect that information is already encoded in the representation rather than capacity added by downstream fine-tuning. The consistent improvements across all four prediction targets therefore suggest that \our{} captures homodimer binding signals at residue-level resolution, providing mechanistic evidence for the complex-level gains observed in \Cref{tab:cofolding-test}.


\begin{figure*}[t]
\centering
\includegraphics[width=.95\linewidth]{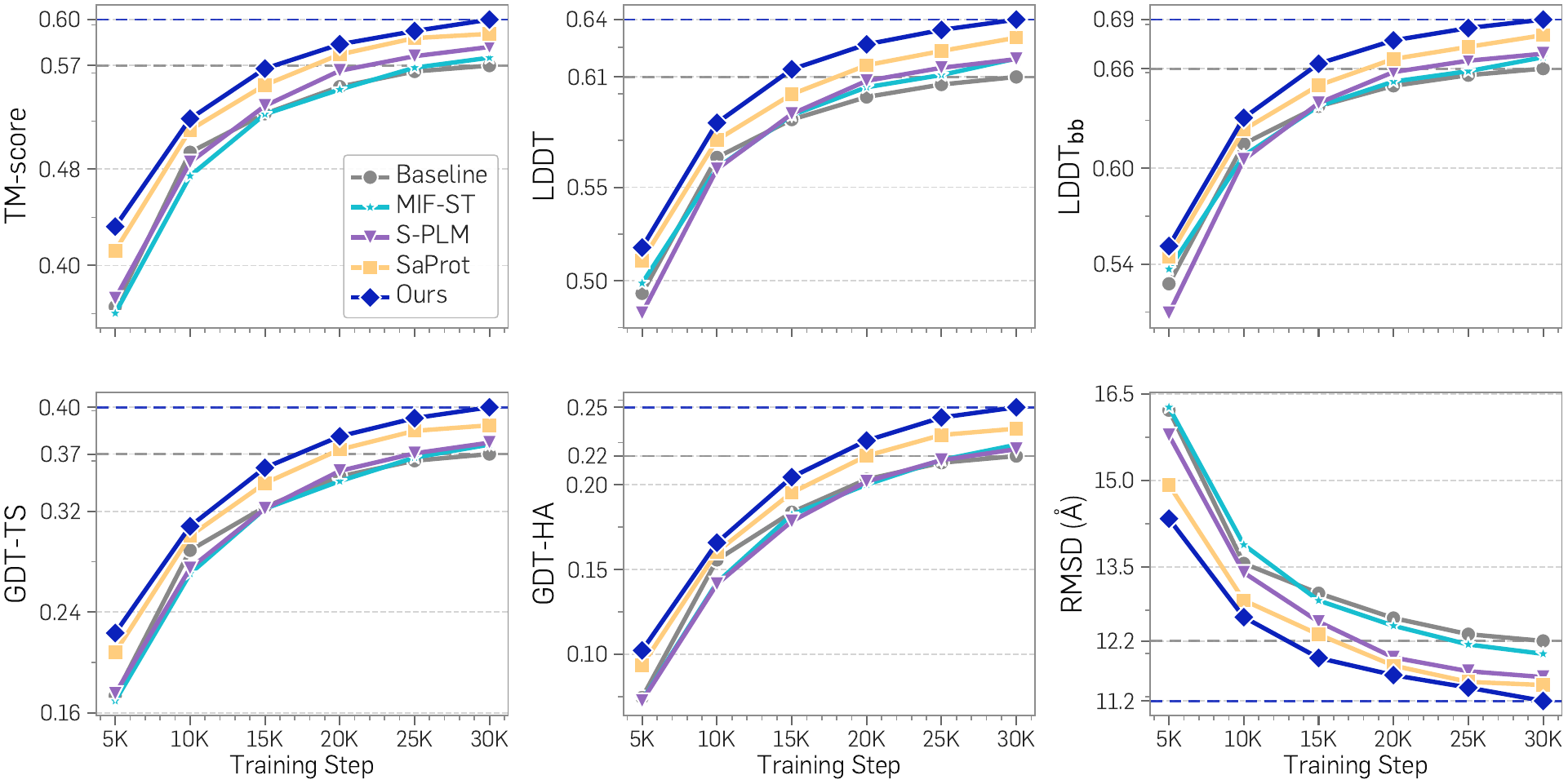}
\caption{\textbf{Acceleration of monomer structure prediction via representation alignment on \bench{}.} We compare the no-REPA baseline against ESM2, SaProt, S-PLM, MIF-ST, and \our as the alignment target. \our provides strongest alignment target for structure prediction model.}\label{fig:apo_repa}
\vspace{-.17in}
\end{figure*}


\begin{table}[t]
\centering
\caption{\textbf{Monomer structure prediction with distillation on \bench{}.} We report test performance of a monomer structure prediction model. \textbf{Bold} denotes the best performance. \our{} provides the strongest alignment target among the evaluated large-scale representation models.}\label{tab:monomer-repa}
\centering
\resizebox{.95\linewidth}{!}{
\setlength{\tabcolsep}{4pt}
\renewcommand{\arraystretch}{0.95}
\begin{tabular}{lc|cccccc}
\toprule
\textbf{REPA target} & \textbf{Params}
& TM-score $\uparrow$ & GDT-TS $\uparrow$ & GDT-HA $\uparrow$
& LDDT $\uparrow$ & LDDT$_\mathrm{bb}$ $\uparrow$ & RMSD (\AA) $\downarrow$ \\
\midrule
Baseline & --   & 0.5422 & 0.3432 & 0.1990 & 0.6208 & 0.6746 & 12.98 \\
SaProt   & 650M & 0.5659 & 0.3677 & 0.2192 & 0.6410 & 0.6947 & 12.26 \\
S-PLM    & 704M & 0.5569 & 0.3559 & 0.2081 & 0.6301 & 0.6838 & 12.40 \\
MIF-ST   & 643M & 0.5491 & 0.3520 & 0.2075 & 0.6310 & 0.6827 & 12.73 \\
\rowcolor{lightblue}
\textcolor{deepblue}{\textbf{Ours}}     & 650M & \textbf{0.5822} & \textbf{0.3858} & \textbf{0.2344} & \textbf{0.6513} & \textbf{0.7051} & \textbf{11.91} \\
\bottomrule
\end{tabular}}
\end{table}

\subsection{Distillation into monomer structure prediction}
\label{sec:exp_folding}
Unlike the co-folding and probing evaluations, this experiment uses pretrained representations not as frozen input features, but as supervisory targets for the structure prediction model. It therefore tests a complementary property of a representation: whether its structural signal is organized in a form that can guide the learning of a generative structure predictor.

\Cref{fig:apo_repa,tab:monomer-repa} report monomer structure prediction results under this representation-alignment setting. \our{} achieves the best performance across all reported metrics, indicating that it provides the most effective supervisory signal for structure prediction training. This suggests that the structural information captured by \our{} is not only decodable by downstream models, but also transferable as a training target for structure-generative learning.


We further evaluate huge-scale baselines in \Cref{tab:huge_apo_repa} of \Cref{appx:huge_apo_repa}. At this scale, performance saturates, with ProstT5 and \our{} achieving comparable results. Although ProstT5 is pretrained for sequence-to-structure translation, \our{} matches its performance without this objective, while also performing well in flexible docking-related evaluations where ProstT5 gives only limited gains. These results indicate that \our{} learns rich representations that support flexible docking and serve as effective alignment targets for monomer structure prediction.



\begin{table}[t]
\centering
\caption{\textbf{Ablations on per-residue homodimer binding properties probing.} MLP probes on monomer protein representations predict properties derived from the holodimer. Ours$_{\text{AF}}$ and Ours$_{\text{MLM}}$ denote variants trained only on AlphaFold structures and with masked language modeling, respectively. Overall, our full model yields strongest performance compared to variants.}
\label{tab:ablation}
\centering
\resizebox{.95\linewidth}{!}{
\setlength{\tabcolsep}{4pt}
\renewcommand{\arraystretch}{0.95}
\begin{tabular}{lc|cc|cc|cc|cc}
\toprule
& & \multicolumn{2}{c|}{\textbf{Binding site}}
& \multicolumn{2}{c|}{\textbf{$\Delta$SASA}}
& \multicolumn{2}{c|}{\textbf{Levy tier}}
& \multicolumn{2}{c}{\textbf{Bond type}} \\
\textbf{Models} & \textbf{Params}
& AUPRC & AUROC 
& Pearson & Spearman 
& Macro F1 & Acc. 
& AUPRC & AUROC \\
\midrule
ESM2  & 150M  & 0.7104 & 0.9066 & 0.6376 & 0.3701 & 0.5701 & 0.7972 & 0.3534 & 0.9542 \\
Ours$_{\text{AF}}$  & 150M & 0.7759 & 0.9314 & 0.7285 & 0.4344 & 0.6458 & 0.8470 & 0.3818 & 0.9576 \\
Ours$_{\text{MLM}}$  & 150M   & 0.7677 & 0.9297 & 0.7309 & 0.4380 & 0.6424 & 0.8471 & 0.3748 & 0.9588 \\
\rowcolor{lightblue}
\textcolor{deepblue}{\textbf{Ours}}    & 150M   & \textbf{0.8156} & \textbf{0.9432} & \textbf{0.7797} & \textbf{0.4662} & \textbf{0.6819} & \textbf{0.8668} & \textbf{0.4184} & \textbf{0.9644} \\
\bottomrule
\end{tabular}}
\end{table}



\subsection{Ablation studies}\label{sec:abblation}

\begin{wraptable}{r}{0.48\linewidth}
\centering
\vspace{-.15in}
\caption{\textbf{Binding-site probing for binder screening.} Probes to predict binding site and contact count, conditioned on binder representation. \our shows strongest performance.}
\label{tab:binder-screening-binding-site}
\resizebox{\linewidth}{!}{
\setlength{\tabcolsep}{3.5pt}
\renewcommand{\arraystretch}{1}
\begin{tabular}{lc|cc|cc}
\toprule
\textbf{Model} & \textbf{Params}
& \multicolumn{2}{c|}{\textbf{Binding site}}
& \multicolumn{2}{c}{\textbf{Contact count}} \\
& & AUPRC & AUROC & Pearson & Spearman \\
\midrule
ESM2    & 3B     & 0.7646 & 0.7743 & 0.4162 & 0.3987 \\
SaProt  & 650M   & 0.7978 & 0.8112 & 0.4996 & 0.4757 \\
Ours & 650M & 0.8184 & 0.8271 & 0.4830 & 0.4700 \\
S-PLM   & 704M   & 0.7541 & 0.7640 & 0.4006 & 0.3903 \\
ProstT5 & 1.2B   & 0.7608 & 0.7688 & 0.4392 & 0.4244 \\
ESM3    & 1.4B   & 0.7472 & 0.7719 & 0.0543 & 0.0551 \\
\rowcolor{lightblue}
\textcolor{deepblue}{\textbf{Ours}}    & 2.8B     & \textbf{0.8332} & \textbf{0.8431} & \textbf{0.5169} & \textbf{0.5020} \\
\bottomrule
\end{tabular}}
\vspace{-.1in}
\end{wraptable}

\textbf{Ablation on pretraining data and objective. } 
~\Cref{tab:ablation} ablates two design choices in \our{} at the 150M scale: pretraining data and pretraining objective. First, Ours${_\text{AF}}$ is trained only on AlphaFold structures, removing ESMAtlas from the pretraining corpus. Next, Ours${_\text{MLM}}$ replaces the generator-corrupted corrective objective with standard masked language modeling. Both variants outperform ESM2-150M, showing that three-view structure-aware pretraining already provides a strong representation, but the full model achieves the best performance. This indicates that both the augmented data and the generator-based corruption improve representation training.


\textbf{Target-conditioned binder-interface probing.}
\Cref{tab:binder-screening-binding-site} evaluates whether protein representations identify target residues involved in heteromeric binding. This setting differs from homodimer probing because the interface is formed with a non-identical binder, and therefore tests whether the representation transfers to asymmetric cross-chain interactions. \our{} shows competitive or stronger performance on both binding-site classification and contact-count regression. These results suggest that \our{} encodes residue-level interface information that transfers beyond homodimeric interactions, supporting its use for target-conditioned binder screening. See the full results and details in~\Cref{tab:binder-screening-binding-site-full} of \Cref{appx:benchmark_heterodimer_balanced}.

\begin{wraptable}{r}{0.45\linewidth}
\vspace{-.15in}
\setlength{\columnsep}{4pt}
\centering
\caption{\textbf{Per-protein function probing on EC and GO.} MLP probing results for EC and GO benchmarks on mean-pooled representations. \our{} shows competitive performance.}
\label{tab:ec-go-probing}
\centering
\resizebox{\linewidth}{!}{
\setlength{\tabcolsep}{3.5pt}
\renewcommand{\arraystretch}{0.95}
\begin{tabular}{lc|c|ccc}
\toprule
& &  & \multicolumn{3}{c}{\textbf{GO}} \\
\textbf{Models} & \textbf{Params}
& \textbf{EC} & \textbf{MF} & \textbf{BP} & \textbf{CC} \\
\hline
\rowcolor{gray!12}\multicolumn{6}{l}{\textit{Small models}} \\
ESM2    & 35M  & 0.888 & 0.618 & 0.456 & 0.521 \\
SaProt  & 35M  & 0.882 & 0.614 & 0.449 & \textbf{0.539} \\
\rowcolor{lightblue}
\textcolor{deepblue}{\textbf{Ours}}    & 35M  & \textbf{0.914} & \textbf{0.640} & \textbf{0.474} & 0.504 \\
\hline
\rowcolor{gray!12}\multicolumn{6}{l}{\textit{Medium models}} \\
ESM2    & 150M & 0.900 & 0.633 & 0.463 & \textbf{0.529} \\
\rowcolor{lightblue}
\textcolor{deepblue}{\textbf{Ours}}    & 150M & \textbf{0.906} & \textbf{0.642} & \textbf{0.467} & 0.528 \\
\hline
\rowcolor{gray!12}\multicolumn{6}{l}{\textit{Large models}} \\
ESM2    & 650M & 0.906 & 0.646 & 0.481 & 0.533 \\
SaProt  & 650M & 0.897 & 0.637 & 0.472 & \textbf{0.555} \\
S-PLM   & 704M & \textbf{0.915} & \textbf{0.660} & 0.480 & 0.531 \\
\rowcolor{lightblue}
\textcolor{deepblue}{\textbf{Ours}}    & 650M & \textbf{0.915} & 0.655 & \textbf{0.488} & 0.511 \\
\hline
\rowcolor{gray!12}\multicolumn{6}{l}{\textit{Huge models}} \\
ESM2    & 3B   & 0.920 & 0.654 & \textbf{0.490} & \textbf{0.532} \\
ESM3    & 1.4B & \textbf{0.926} & 0.648 & 0.487 & 0.516 \\
\rowcolor{lightblue}
\textcolor{deepblue}{\textbf{Ours}}    & 2.8B   & 0.923 & \textbf{0.660} & \textbf{0.490} & 0.508 \\
\bottomrule
\end{tabular}}
\end{wraptable}

\textbf{Conventional benchmarks.} We also evaluate \our{} on conventional benchmarks, namely Enzyme Commission and Gene Ontology~\citep{gligorijevic2021structure}. We use a two-layer MLP probing setup. We describe the detailed hyperparameters in \Cref{appx:conventional_benchmark}. As shown in \Cref{tab:ec-go-probing}, \our{} remains competitive with the strongest baselines on these benchmarks, while sequence-only ESM2-3B performs close to structure-aware models.

\begin{table}[t]
\centering
\caption{{\textbf{Homodimer structure prediction on RCSB structures deposited after June 1, 2023.} We evaluate the co-folding models in \Cref{tab:cofolding-test} on real-world homodimers. \textbf{Bold} denotes the best performance. The relative gains of \our{} are preserved in predicting real-world structures.}}
\label{tab:cofolding-encoder}
\makebox[\textwidth][c]{%
\resizebox{0.9\textwidth}{!}{%
\setlength{\tabcolsep}{4pt}
\renewcommand{\arraystretch}{0.88}
\begin{tabular}{lc|ccc|ccc}
\toprule
& & \multicolumn{3}{c|}{\textbf{Max seq. id. $<$ 90\% (N$=$252)}}
& \multicolumn{3}{c}{\textbf{Max seq. id. $<$ 40\% (N$=$86)}} \\
\textbf{Models} & \textbf{Params} 
& DockQ $\uparrow$ & Medium\% $\uparrow$ & High\% $\uparrow$
& DockQ $\uparrow$ & Medium\% $\uparrow$ & High\% $\uparrow$ \\
\midrule
S-PLM    & 704M & 0.275 & 29.8 & 4.0 & 0.185 & 17.4 & 3.5 \\
ProstT5  & 1.2B & 0.297 & 33.3 & 5.6 & 0.193 & 18.6 & 2.3 \\
ESM2     & 3B   & 0.296 & 32.9 & 6.0 & 0.197 & 18.6 & 2.3 \\
SaProt   & 650M & 0.320 & 36.5 & 7.1 & 0.210 & 19.8 & 4.7 \\
ESM3     & 1.4B & 0.322 & 37.7 & 8.7 & 0.195 & 19.8 & 3.5 \\
\rowcolor{lightblue}
\textcolor{deepblue}{\textbf{Ours}} & 2.8B
& \textbf{0.342} & \textbf{39.7} & \textbf{13.5}
& \textbf{0.228} & \textbf{20.9} & \textbf{8.1} \\
\bottomrule
\end{tabular}%
}%
}
\vspace{-.2in}
\end{table}

\textbf{Performance on real-world protein structures.}
We further evaluate whether the co-folding results transfer beyond the AFDB-derived benchmark. \Cref{tab:cofolding-encoder} reports homodimer structure prediction performance on RCSB structures deposited after June 1, 2023 \citep{burley2021rcsb}. Since our pretraining and co-folding training use predicted AFDB structures rather than experimentally resolved structures, we mainly use loss-based filtering and additionally report results under the conventional cutoff. The relative trend observed on \bench{} is preserved on these real-world structures: \our{} consistently outperforms the baselines in average DockQ score and in medium- and high-quality success rates (DockQ$\geq0.49$ and DockQ$\geq0.8$).



\section{Conclusion}
\label{sec:conclusion}
In this paper, we presented \our{}, a structure-aware representation model that incorporates amino-acid identity, backbone geometry, and full-atom residue geometry, together with \bench{}, a benchmark for evaluating whether pretrained protein representations provide useful geometric signals for structure-predictive modeling. On \bench{}, structure-aware representations outperform sequence-only protein representations, and \our{} further improves over prior structure-aware baselines. These results highlight the utility of structure-aware protein representations for structure-predictive modeling, with finer geometric detail further improving their utility in these tasks.

\textbf{Limitations. }Following prior structure-aware representation learning studies that use predicted structures~\citep{su2024saprot,heinzinger2024bilingual,wang2025s}, \our{} and \bench{} use high-confidence AFDB structures for million-scale, uniformly processed benchmarking, while broader validation on experimentally resolved structures remains future work despite similar initial trends on real PDB complexes (\Cref{tab:cofolding-encoder}).



\bibliographystyle{unsrt}
\bibliography{references}

@article{jumper2021highly,
  title={Highly accurate protein structure prediction with AlphaFold},
  author={Jumper, John and Evans, Richard and Pritzel, Alexander and Green, Tim and Figurnov, Michael and Ronneberger, Olaf and Tunyasuvunakool, Kathryn and Bates, Russ and {\v{Z}}{\'\i}dek, Augustin and Potapenko, Anna and others},
  journal={nature},
  volume={596},
  number={7873},
  pages={583--589},
  year={2021},
  publisher={Nature Publishing Group UK London}
}

@article{sereyjol2026test,
  title={Test-Time Conditioning with Representation-Aligned Visual Features},
  author={Sereyjol-Garros, Nicolas and Kirby, Ellington and Letzelter, Victor and Besnier, Victor and Samet, Nermin},
  journal={arXiv preprint arXiv:2602.03753},
  year={2026}
}

@article{dssp,
  title   = {Dictionary of protein secondary structure: pattern recognition of hydrogen-bonded and geometrical features},
  author  = {Kabsch, Wolfgang and Sander, Christian},
  journal = {Biopolymers},
  volume  = {22},
  number  = {12},
  pages   = {2577--2637},
  year    = {1983},
  doi     = {10.1002/bip.360221211}
}

@inproceedings{leng2025repa,
  title={Repa-e: Unlocking vae for end-to-end tuning of latent diffusion transformers},
  author={Leng, Xingjian and Singh, Jaskirat and Hou, Yunzhong and Xing, Zhenchang and Xie, Saining and Zheng, Liang},
  booktitle={Proceedings of the IEEE/CVF International Conference on Computer Vision},
  pages={18262--18272},
  year={2025}
}

@article{ye2023ip-adapter,
  title={IP-Adapter: Text Compatible Image Prompt Adapter for Text-to-Image Diffusion Models},
  author={Ye, Hu and Zhang, Jun and Liu, Sibo and Han, Xiao and Yang, Wei},
  booktitle={arXiv preprint arxiv:2308.06721},
  year={2023}
}

@inproceedings{pipref50k,
  title     = {Learning to Design Protein-Protein Interactions with Enhanced Generalization},
  author    = {Bushuiev, Anton and Bushuiev, Roman and Kouba, Petr and Filkin, Anatolii and Gabrielova, Marketa and Gabriel, Michal and Sedlar, Jiri and Pluskal, Tomas and Damborsky, Jiri and Mazurenko, Stanislav and Sivic, Josef},
  booktitle = {The Twelfth International Conference on Learning Representations},
  year      = {2024},
  url       = {https://openreview.net/forum?id=xcMmebCT7s},
  eprint    = {2310.18515},
  archivePrefix = {arXiv}
}

@article{burley2021rcsb,
  title={RCSB Protein Data Bank: powerful new tools for exploring 3D structures of biological macromolecules for basic and applied research and education in fundamental biology, biomedicine, biotechnology, bioengineering and energy sciences},
  author={Burley, Stephen K and Bhikadiya, Charmi and Bi, Chunxiao and Bittrich, Sebastian and Chen, Li and Crichlow, Gregg V and Christie, Cole H and Dalenberg, Kenneth and Di Costanzo, Luigi and Duarte, Jose M and others},
  journal={Nucleic acids research},
  volume={49},
  number={D1},
  pages={D437--D451},
  year={2021},
  publisher={Oxford University Press}
}

@article{varadi2022alphafold,
  title={AlphaFold Protein Structure Database: massively expanding the structural coverage of protein-sequence space with high-accuracy models},
  author={Varadi, Mihaly and Anyango, Stephen and Deshpande, Mandar and Nair, Sreenath and Natassia, Cindy and Yordanova, Galabina and Yuan, David and Stroe, Oana and Wood, Gemma and Laydon, Agata and others},
  journal={Nucleic acids research},
  volume={50},
  number={D1},
  pages={D439--D444},
  year={2022},
  publisher={Oxford University Press}
}

@inproceedings{
su2024saprot,
title={SaProt: Protein Language Modeling with Structure-aware Vocabulary},
author={Jin Su and Chenchen Han and Yuyang Zhou and Junjie Shan and Xibin Zhou and Fajie Yuan},
booktitle={The Twelfth International Conference on Learning Representations},
year={2024},
url={https://openreview.net/forum?id=6MRm3G4NiU}
}

@article{hayes2025simulating,
  title={Simulating 500 million years of evolution with a language model},
  author={Hayes, Thomas and Rao, Roshan and Akin, Halil and Sofroniew, Nicholas J and Oktay, Deniz and Lin, Zeming and Verkuil, Robert and Tran, Vincent Q and Deaton, Jonathan and Wiggert, Marius and others},
  journal={Science},
  volume={387},
  number={6736},
  pages={850--858},
  year={2025},
  publisher={American Association for the Advancement of Science}
}

@article{heinzinger2024bilingual,
    author = {Heinzinger, Michael and Weissenow, Konstantin and Sanchez, Joaquin Gomez and Henkel, Adrian and Mirdita, Milot and Steinegger, Martin and Rost, Burkhard},
    title = {Bilingual language model for protein sequence and structure},
    journal = {NAR Genomics and Bioinformatics},
    volume = {6},
    number = {4},
    pages = {lqae150},
    year = {2024},
    month = {12},
    issn = {2631-9268},
    doi = {10.1093/nargab/lqae150},
    url = {https://doi.org/10.1093/nargab/lqae150},
    eprint = {https://academic.oup.com/nargab/article-pdf/6/4/lqae150/60777547/lqae150.pdf},
}

@article{steinegger2017mmseqs2,
  title={MMseqs2 enables sensitive protein sequence searching for the analysis of massive data sets},
  author={Steinegger, Martin and S{\"o}ding, Johannes},
  journal={Nature biotechnology},
  volume={35},
  number={11},
  pages={1026--1028},
  year={2017},
  publisher={Nature Publishing Group US New York}
}

@article{ec_and_go,
  title     = {Structure-based protein function prediction using graph convolutional networks},
  author    = {Gligorijevi{\'c}, Vladimir and Renfrew, P. Douglas and Kosciolek, Tomasz and Leman, Julia Koehler and Berenberg, Daniel and Vatanen, Tommi and Chandler, Chris and Taylor, Bryn C. and Fisk, Ian M. and Vlamakis, Hera and Xavier, Ramnik J. and Knight, Rob and Cho, Kyunghyun and Bonneau, Richard},
  journal   = {Nature Communications},
  volume    = {12},
  number    = {1},
  pages     = {3168},
  year      = {2021},
  doi       = {10.1038/s41467-021-23303-9},
  url       = {https://doi.org/10.1038/s41467-021-23303-9}
}

@article{varadi2024alphafold,
    author = {Varadi, Mihaly and Bertoni, Damian and Magana, Paulyna and Paramval, Urmila and Pidruchna, Ivanna and Radhakrishnan, Malarvizhi and Tsenkov, Maxim and Nair, Sreenath and Mirdita, Milot and Yeo, Jingi and Kovalevskiy, Oleg and Tunyasuvunakool, Kathryn and Laydon, Agata and Žídek, Augustin and Tomlinson, Hamish and Hariharan, Dhavanthi and Abrahamson, Josh and Green, Tim and Jumper, John and Birney, Ewan and Steinegger, Martin and Hassabis, Demis and Velankar, Sameer},
    title = {AlphaFold Protein Structure Database in 2024: providing structure coverage for over 214 million protein sequences},
    journal = {Nucleic Acids Research},
    volume = {52},
    number = {D1},
    pages = {D368-D375},
    year = {2024},
    month = {01},
    issn = {0305-1048},
    doi = {10.1093/nar/gkad1011},
    url = {https://doi.org/10.1093/nar/gkad1011},
    eprint = {https://academic.oup.com/nar/article-pdf/52/D1/D368/55039845/gkad1011.pdf},
}

@article{li2024return,
  title={Return of unconditional generation: A self-supervised representation generation method},
  author={Li, Tianhong and Katabi, Dina and He, Kaiming},
  journal={Advances in Neural Information Processing Systems},
  volume={37},
  pages={125441--125468},
  year={2024}
}

@article{wang2025s,
  title={S-PLM: structure-aware protein language model via contrastive learning between sequence and structure},
  author={Wang, Duolin and Pourmirzaei, Mahdi and Abbas, Usman L and Zeng, Shuai and Manshour, Negin and Esmaili, Farzaneh and Poudel, Biplab and Jiang, Yuexu and Shao, Qing and Chen, Jin and others},
  journal={Advanced Science},
  volume={12},
  number={5},
  pages={2404212},
  year={2025},
  publisher={Wiley Online Library}
}

@article{
lin2023evolutionary,
author = {Zeming Lin  and Halil Akin  and Roshan Rao  and Brian Hie  and Zhongkai Zhu  and Wenting Lu  and Nikita Smetanin  and Robert Verkuil  and Ori Kabeli  and Yaniv Shmueli  and Allan dos Santos Costa  and Maryam Fazel-Zarandi  and Tom Sercu  and Salvatore Candido  and Alexander Rives },
title = {Evolutionary-scale prediction of atomic-level protein structure with a language model},
journal = {Science},
volume = {379},
number = {6637},
pages = {1123-1130},
year = {2023},
doi = {10.1126/science.ade2574},
URL = {https://www.science.org/doi/abs/10.1126/science.ade2574},
eprint = {https://www.science.org/doi/pdf/10.1126/science.ade2574}}

@article{vankempen2024fast,
  title={Fast and accurate protein structure search with Foldseek},
  author={Van Kempen, Michel and Kim, Stephanie S and Tumescheit, Charlotte and Mirdita, Milot and Lee, Jeongjae and Gilchrist, Cameron LM and S{\"o}ding, Johannes and Steinegger, Martin},
  journal={Nature biotechnology},
  volume={42},
  number={2},
  pages={243--246},
  year={2024},
  publisher={Nature Publishing Group US New York}
}

@article{van2017neural,
  title={Neural discrete representation learning},
  author={Van Den Oord, Aaron and Vinyals, Oriol and others},
  journal={Advances in neural information processing systems},
  volume={30},
  year={2017}
}

@inproceedings{peebles2023scalable,
  title={Scalable diffusion models with transformers},
  author={Peebles, William and Xie, Saining},
  booktitle={Proceedings of the IEEE/CVF international conference on computer vision},
  pages={4195--4205},
  year={2023}
}

@inproceedings{clark2020electra,
  title = {{ELECTRA}: Pre-training Text Encoders as Discriminators Rather Than Generators},
  author = {Kevin Clark and Minh-Thang Luong and Quoc V. Le and Christopher D. Manning},
  booktitle = {ICLR},
  year = {2020},
  url = {https://openreview.net/pdf?id=r1xMH1BtvB}
}

@article{xu2020mc,
  title={Mc-bert: Efficient language pre-training via a meta controller},
  author={Xu, Zhenhui and Gong, Linyuan and Ke, Guolin and He, Di and Zheng, Shuxin and Wang, Liwei and Bian, Jiang and Liu, Tie-Yan},
  journal={arXiv preprint arXiv:2006.05744},
  year={2020}
}

@inproceedings{
yu2025representation,
title={Representation Alignment for Generation: Training Diffusion Transformers Is Easier Than You Think},
author={Sihyun Yu and Sangkyung Kwak and Huiwon Jang and Jongheon Jeong and Jonathan Huang and Jinwoo Shin and Saining Xie},
booktitle={The Thirteenth International Conference on Learning Representations},
year={2025},
url={https://openreview.net/forum?id=DJSZGGZYVi}
}

@article{wohlwend2025boltz,
  title={Boltz-1 democratizing biomolecular interaction modeling},
  author={Wohlwend, Jeremy and Corso, Gabriele and Passaro, Saro and Getz, Noah and Reveiz, Mateo and Leidal, Ken and Swiderski, Wojtek and Atkinson, Liam and Portnoi, Tally and Chinn, Itamar and others},
  journal={BioRxiv},
  pages={2024--11},
  year={2025}
}

@article{passaro2025boltz2,
  title={Boltz-2: Towards accurate and efficient binding affinity prediction},
  author={Passaro, Saro and Corso, Gabriele and Wohlwend, Jeremy and Reveiz, Mateo and Thaler, Stephan and Somnath, Vignesh Ram and Getz, Noah and Portnoi, Tally and Roy, Julien and Stark, Hannes and others},
  journal={BioRxiv},
  year={2025}
}

@article{abramson2024accurate,
  title={Accurate structure prediction of biomolecular interactions with AlphaFold 3},
  author={Abramson, Josh and Adler, Jonas and Dunger, Jack and Evans, Richard and Green, Tim and Pritzel, Alexander and Ronneberger, Olaf and Willmore, Lindsay and Ballard, Andrew J and Bambrick, Joshua and others},
  journal={Nature},
  volume={630},
  number={8016},
  pages={493--500},
  year={2024},
  publisher={Nature Publishing Group UK London}
}

@article{evans2021protein,
  title={Protein complex prediction with AlphaFold-Multimer},
  author={Evans, Richard and O’neill, Michael and Pritzel, Alexander and Antropova, Natasha and Senior, Andrew and Green, Tim and {\v{Z}}{\'\i}dek, Augustin and Bates, Russ and Blackwell, Sam and Yim, Jason and others},
  journal={biorxiv},
  pages={2021--10},
  year={2021},
  publisher={Cold Spring Harbor Laboratory}
}

@article{gligorijevic2021structure,
  title={Structure-based protein function prediction using graph convolutional networks},
  author={Gligorijevi{\'c}, Vladimir and Renfrew, P Douglas and Kosciolek, Tomasz and Leman, Julia Koehler and Berenberg, Daniel and Vatanen, Tommi and Chandler, Chris and Taylor, Bryn C and Fisk, Ian M and Vlamakis, Hera and others},
  journal={Nature communications},
  volume={12},
  number={1},
  pages={3168},
  year={2021},
  publisher={Nature Publishing Group UK London}
}

@article{yang2023masked,
  title={Masked inverse folding with sequence transfer for protein representation learning},
  author={Yang, Kevin K and Zanichelli, Niccol{\`o} and Yeh, Hugh},
  journal={Protein Engineering, Design and Selection},
  volume={36},
  pages={gzad015},
  year={2023},
  publisher={Oxford University Press}
}

@inproceedings{yuan2025protein,
  title={Protein Structure Tokenization: Benchmarking and New Recipe},
  author={Yuan, Xinyu and Wang, Zichen and Collins, Marcus D and Rangwala, Huzefa},
  booktitle={International Conference on Machine Learning},
  pages={73645--73670},
  year={2025},
  organization={PMLR}
}

@article{lee1971interpretation,
  title={The interpretation of protein structures: estimation of static accessibility},
  author={Lee, Byungkook and Richards, Frederic M},
  journal={Journal of molecular biology},
  volume={55},
  number={3},
  pages={379--IN4},
  year={1971},
  publisher={Elsevier}
}

@article{shrake1973environment,
  title={Environment and exposure to solvent of protein atoms. Lysozyme and insulin},
  author={Shrake, Andrew and Rupley, John A},
  journal={Journal of molecular biology},
  volume={79},
  number={2},
  pages={351--371},
  year={1973},
  publisher={Elsevier}
}

@article{levy2010simple,
  title={A simple definition of structural regions in proteins and its use in analyzing interface evolution},
  author={Levy, Emmanuel D},
  journal={Journal of molecular biology},
  volume={403},
  number={4},
  pages={660--670},
  year={2010},
  publisher={Elsevier}
}

@article{salentin2015plip,
  title={PLIP: fully automated protein--ligand interaction profiler},
  author={Salentin, Sebastian and Schreiber, Sven and Haupt, V Joachim and Adasme, Melissa F and Schroeder, Michael},
  journal={Nucleic acids research},
  volume={43},
  number={W1},
  pages={W443--W447},
  year={2015},
  publisher={Oxford University Press}
}

@article{jubb2017arpeggio,
  title={Arpeggio: a web server for calculating and visualising interatomic interactions in protein structures},
  author={Jubb, Harry C and Higueruelo, Alicia P and Ochoa-Monta{\~n}o, Bernardo and Pitt, Will R and Ascher, David B and Blundell, Tom L},
  journal={Journal of molecular biology},
  volume={429},
  number={3},
  pages={365--371},
  year={2017},
  publisher={Elsevier}
}

@inproceedings{
wang_simplefold_2025,
title={SimpleFold: Folding Proteins is Simpler than You Think},
author={Yuyang Wang and Jiarui Lu and Navdeep Jaitly and Joshua M. Susskind and Miguel {\'A}ngel Bautista},
booktitle={The Fourteenth International Conference on Learning Representations},
year={2026},
url={https://openreview.net/forum?id=0j0MmK7EMA}
}

@article{cock2009biopython,
  title = {Biopython: freely available Python tools for computational molecular biology and bioinformatics},
  author = {Cock, Peter J. A. and Antao, Tiago and Chang, Jeffrey T. and Chapman, Brad A. and Cox, Cymon J. and Dalke, Andrew and Friedberg, Iddo and Hamelryck, Thomas and Kauff, Frank and Wilczynski, Bartek and de Hoon, Michiel J. L.},
  journal = {Bioinformatics},
  volume = {25},
  number = {11},
  pages = {1422--1423},
  year = {2009},
  doi = {10.1093/bioinformatics/btp163}
}

@article{han2026alphafold,
  title={AlphaFold Database expands to proteome-scale quaternary structures},
  author={Han, Yewon and Tsenkov, Maxim I and Venanzi, Niccolo AE and Bertoni, Damian and Cha, Sooyoung and Chacon, Alejandro and Dietrich, Nick and Fomitchev, Boris and Goldtzvik, Yonathan and Hsu, Darren and others},
  journal={bioRxiv},
  pages={2026--03},
  year={2026},
  publisher={Cold Spring Harbor Laboratory}
}

@article{mitchell2020mgnify,
  title={MGnify: the microbiome analysis resource in 2020},
  author={Mitchell, Alex L and Almeida, Alexandre and Beracochea, Martin and Boland, Miguel and Burgin, Josephine and Cochrane, Guy and Crusoe, Michael R and Kale, Varsha and Potter, Simon C and Richardson, Lorna J and others},
  journal={Nucleic acids research},
  volume={48},
  number={D1},
  pages={D570--D578},
  year={2020},
  publisher={Oxford University Press}
}

@misc{uniprot2023uniprot,
  title={UniProt: the universal protein knowledgebase in 2023},
  author={The UniProt Consortium},
  journal={Nucleic acids research},
  volume={51},
  number={D1},
  pages={D523--D531},
  year={2023},
  publisher={Oxford University Press}
}

\appendix
\newpage

\section{Details in \our{}}\label[appendix]{appx:method}

\subsection{Full-atom tokenization}\label[appendix]{appx:fullatom_token}

For each residue, the tokenizer computes heavy-atom geometry features from Atom37 coordinates. These features include (i) atomic displacements relative to $\mathrm{C}\alpha$ in a per-residue canonical frame, which gives SE(3) invariance, (ii) distances from $\mathrm{C}\alpha$ to each heavy atom, (iii) binned backbone bond angles, and (iv) binned side-chain torsions $\chi_1,\ldots,\chi_4$. These features initialize a per-atom single representation and a $37 \times 37$ inter-atomic pair representation. The two representations are jointly refined by $N$ stacked Pairformer-style layers. We pool the resulting $37$ atom embeddings within each residue to obtain a continuous embedding $z$, and assign $z$ to its nearest codebook entry $q_k$ from a learned codebook of size $V$.

\textbf{Training objective.}
We pretrain the tokenizer on AlphaFoldDB monomer structures with
\begin{equation}
\mathcal{L}
=
\mathcal{L}_{\mathrm{VQ\mbox{-}VAE}}
+
\mathcal{L}_{\mathrm{pair}}
+
\mathcal{L}_{\chi},
\end{equation}
where $\mathcal{L}_{\mathrm{VQ\mbox{-}VAE}}$ is the standard VQ-VAE objective, $\mathcal{L}_{\mathrm{pair}}$ is a clamped pairwise inter-atomic distance loss, and $\mathcal{L}_{\chi}$ is a per-bin cross-entropy loss for side-chain torsion prediction.

\begin{wrapfigure}{r}{0.35\columnwidth}
\vspace{-0.8em}
\centering
\includegraphics[width=0.88\linewidth]{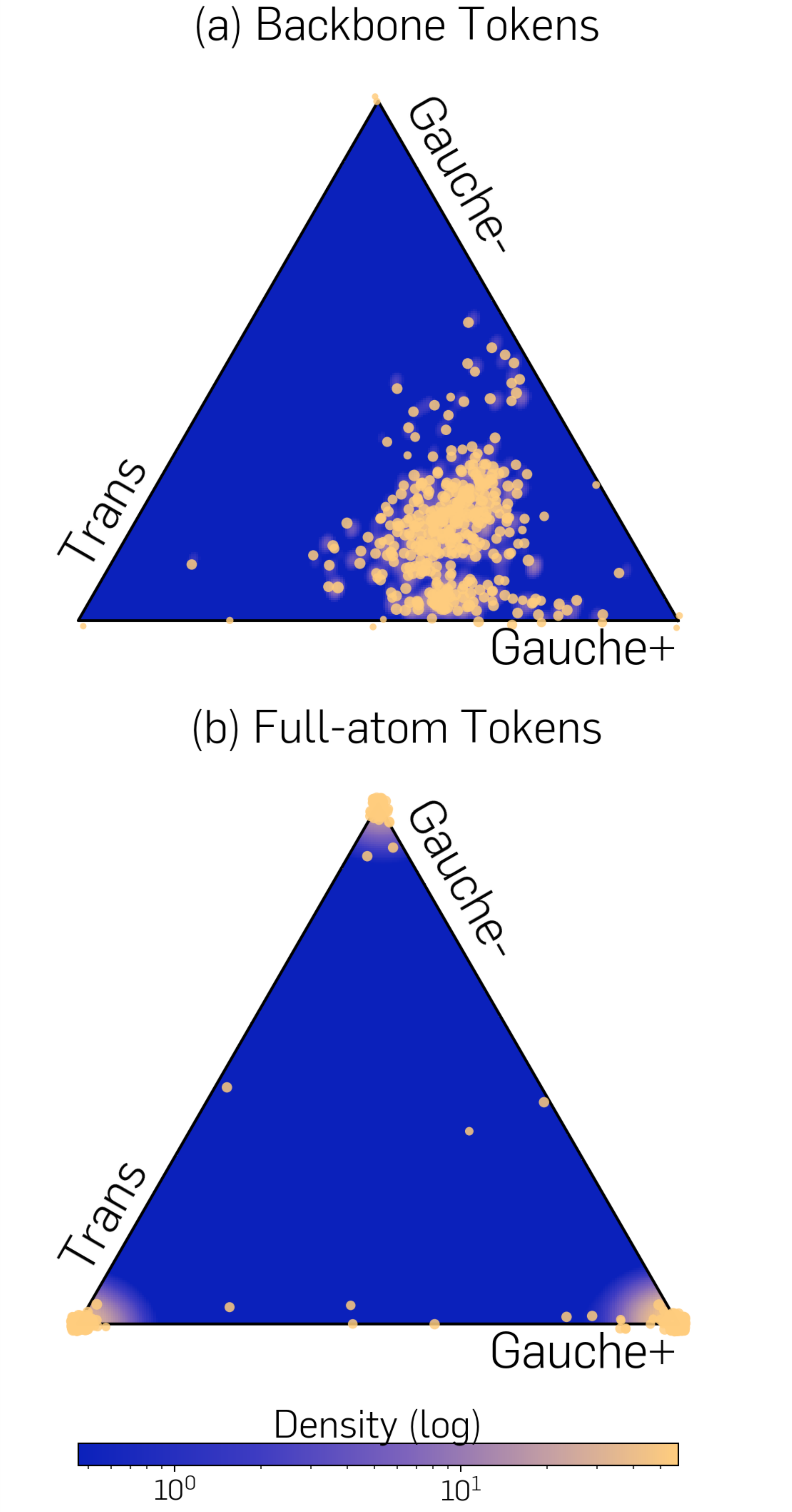}
\vspace{-.05in}
\caption{\textbf{Tokens vs.\ sidechain rotamer.}
Density of codes in the $\chi_1$ simplex. (a)~Backbone tokens. (b)~Full-atom tokens.}
\label{fig:tokenizer_comparison}
\vspace{-0.2in}
\end{wrapfigure}

\textbf{Hyperparameters.}
The tokenizer uses single width $256$, pair width $128$, and $N=6$ Pairformer-style layers with $8$ attention heads. The output embedding dimension is $256$. The codebook contains $V=512$ entries, uses EMA updates with decay $0.99$, and uses entropy regularization with weight $0.1$. Backbone angles and side-chain torsions are discretized into $21$ bins over $[-\pi,\pi]$. The decoder is a $4$-layer Transformer with hidden size $256$ and $8$ attention heads. All loss terms have weight $1.0$. We train with AdamW using learning rate $5{\times}10^{-4}$, $\beta=(0.9,0.99)$, and weight decay $0.01$.

\textbf{Ablations.}
We compare whether the backbone and full-atom tokenizers capture side-chain information by measuring their sensitivity to side-chain rotamers (\Cref{fig:tokenizer_comparison}). For each token, we compute its distribution over the three canonical $\chi_1$ rotamers: \textbf{Trans} ($\approx 180^{\circ}$), \textbf{Gauche$^{+}$} ($\approx +60^{\circ}$), and \textbf{Gauche$^{-}$} ($\approx -60^{\circ}$). We then plot each token in a triangle whose vertices correspond to the three rotamer states. Tokens near a vertex are associated with a specific rotamer, while tokens near the center mix multiple rotamers.

Backbone tokens are broadly spread near the center, indicating that they do not separate side-chain rotamer states well. In contrast, full-atom tokens are concentrated near the vertices, showing that they capture side-chain rotamer geometry. This result confirms that full-atom tokens provide side-chain information that is largely absent from backbone tokens, making the two tokenizers complementary.


\subsection{Corrective pretraining with generator-corrupted views}\label[appendix]{appx:pretrain}

Let
$\boldsymbol{x}=(\boldsymbol{x}^{\mathrm{aa}},\boldsymbol{x}^{\mathrm{bb}},\boldsymbol{x}^{\mathrm{fa}})$
denote the token sequences for a protein of length $L$, where
$\boldsymbol{x}^{\mathrm{aa}}$, $\boldsymbol{x}^{\mathrm{bb}}$, and
$\boldsymbol{x}^{\mathrm{fa}}$ are the amino-acid, backbone, and full-atom token sequences, respectively. We pretrain \our{} with a corrective token-recovery objective. The objective follows the generator-discriminator structure of ELECTRA~\citep{clark2020electra}. A lightweight generator $\mathcal{G}$ replaces selected tokens with plausible alternatives, and the representation model $\mathcal{E}$ recovers the original tokens from the corrupted input.

\textbf{Generator-based tri-token augmentation.}
We use generator-based replacement as an augmentation scheme for tri-token protein pretraining. Instead of masking tokens and asking the model to reconstruct them from explicit mask symbols, we first use $\mathcal{G}$ to replace selected tokens with plausible in-distribution alternatives. We sample replacement positions independently for the three views.

Let
$\boldsymbol{m}^{v} \subset \{1,\ldots,L\}$ denote the positions selected for replacement. We first form a masked input $\boldsymbol{x}^{\mathrm{masked}}$ by replacing $x_i^v$ with $[\mathrm{MASK}]$ for all $i \in \boldsymbol{m}^v$. The generator predicts a token distribution
$p_{\mathcal{G}}^v(\cdot \mid \boldsymbol{x}^{\mathrm{masked}})_i$
at each selected residue $i$ and samples a replacement token:
\begin{equation}
\hat{x}_i^v
\sim
p_{\mathcal{G}}^v(\cdot \mid \boldsymbol{x}^{\mathrm{masked}})_i,
\qquad
i \in \boldsymbol{m}^{v}.
\end{equation}
We then replace the original tokens at the selected positions with the sampled tokens to obtain the corrupted tri-token input $\boldsymbol{x}^{\mathrm{corrupt}}$.

\textbf{Training objective.}
The generator is trained with masked-token cross-entropy:
\begin{equation}
\mathcal{L}_{\mathrm{MLM}}
=
\mathbb{E}
\left[
\sum_{v}
\sum_{i \in \boldsymbol{m}^{v}}
\mathcal{L}_{\mathrm{CE}}
\left(
p_{\mathcal{G}}^{v}(\boldsymbol{x}^{\mathrm{masked}})_i,
x_i^{v}
\right)
\right],
\end{equation}
where  $\mathcal{L}_{\mathrm{CE}}$ denotes the cross-entropy loss over the predicted token distribution.

The representation model is trained to recover the original tokens from the corrupted tri-token input:
\begin{equation}
\mathcal{L}_{\mathrm{rec}}
=
\mathbb{E}
\left[
\sum_{v}
\sum_{i=1}^{L}
\mathcal{L}_{\mathrm{CE}}
\left(
p_{\mathcal{E}}^{v}(\boldsymbol{x}^{\mathrm{corrupt}})_i,
x_i^{v}
\right)
\right],
\end{equation}
where $p_{\mathcal{E}}^v(\cdot \mid \boldsymbol{x}^{\mathrm{corrupt}})_i$
is its predicted token distribution at residue $i$ and view $v$. Unlike the generator loss, the recovery loss is applied to all residue positions and all views.

\textbf{Architecture.}
The generator $\mathcal{G}$ and the representation model $\mathcal{E}$ are Transformer-based sequence models over residue-level tri-token inputs. For each residue $i$, the amino-acid, backbone, and full-atom tokens
$(x_i^{\mathrm{aa}}, x_i^{\mathrm{bb}}, x_i^{\mathrm{fa}})$
are embedded with view-specific embedding tables
$e^{\mathrm{aa}}$, $e^{\mathrm{bb}}$, and $e^{\mathrm{fa}}$. The three embeddings are concatenated along the feature dimension and projected to the Transformer hidden dimension with a fuse layer:
\begin{equation}
h_i^{(0)}
=
W_{\mathrm{fuse}}
\left[
e^{\mathrm{aa}}(x_i^{\mathrm{aa}});
e^{\mathrm{bb}}(x_i^{\mathrm{bb}});
e^{\mathrm{fa}}(x_i^{\mathrm{fa}})
\right].
\end{equation}
The sequence of fused residue representations
$(h_1^{(0)},\ldots,h_L^{(0)})$
is then processed by a Transformer stack with rotary positional information. Note that the generator is smaller than the representation model. After pretraining, we discard the generator and all token-recovery heads. We use the hidden states from the Transformer stack of $\mathcal{E}$ as transferable protein representations.

\textbf{Hyperparameters.} 
We pretrain four model scales, 35M, 150M, 650M, and 2.8B, using the configurations in \Cref{tab:electra-hparams}. Across all scales, the generator $\mathcal{G}$ uses an 8-layer Transformer encoder and a 2-layer Transformer decoder, and is discarded after pretraining. The representation model $\mathcal{E}$ follows the scale-specific Transformer configuration in \Cref{tab:electra-hparams}.

All models are pretrained for 500K steps with a crop size of 512 residues. We use NVIDIA B200 nodes with 8 GPUs per node. The 35M, 150M, 650M, and 2.8B models are trained on 1, 2, 4, and 4 nodes, respectively, with per-GPU batch sizes of 128, 64, 32, and 32. This gives an effective batch size of 1024 protein sequences for all scales. We independently mask each token view with probability 0.6. Training uses AdamW with peak learning rate $4\times 10^{-4}$, weight decay 0.01, $\beta_1=0.9$, $\beta_2=0.95$, gradient clipping at 1.0, and \texttt{bf16-mixed} precision.

\begin{table}[t]
\centering
\caption{ELECTRA model configuration across the four scales.}
\label{tab:electra-hparams}
\small
\setlength{\tabcolsep}{6pt}
\begin{tabular}{lcccc}
\toprule
                                  & 35M    & 150M   & 650M    & 2.8B     \\
\midrule
\quad embedding dim               & 480    & 640    & 1280    & 2560   \\
\quad encoder depth               & 10     & 28     & 30      & 33     \\
\quad decoder depth               & 2      & 2      & 3       & 3      \\
\quad attention heads             & 20     & 20     & 20      & 40     \\
\bottomrule
\end{tabular}
\end{table}

\newpage

\section{Details in \bench}\label[appendix]{appx:benchmark}

\subsection{Dataset curation}\label[appendix]{appx:benchmark_curation}

We use $1.8$M high-confidence homodimer complexes released in the March 2026 AlphaFold Protein Structure Database (AFDB) complex expansion~\citep{evans2021protein,varadi2024alphafold,han2026alphafold}. For each dimer, we obtain the corresponding single-chain apo monomer prediction from AFDB, keyed by UniProt accession. Samples without an AFDB monomer prediction are removed, leaving 1{,}719{,}497 apo-holo pairs. We split the benchmark by sequence clusters using MMseqs2~\citep{steinegger2017mmseqs2} \texttt{easy-cluster} with a 50\% sequence-identity threshold. This yields 404{,}961 cluster representatives, from which we select 400 validation and 1{,}000 test sequences. We construct the initial training split from the remaining representatives, then remove training sequences that share more than 50\% sequence identity with any validation or test sequence. After this filtering step, the training split contains 390{,}861 samples.

\subsection{Homodimer structure prediction}\label[appendix]{appx:homodimer_prediction}
To construct a co-folding model for benchmark purposes, we modify SimpleFold~\citep{wang_simplefold_2025}, an open Transformer-based single-protein folding model, into a homodimer co-folding model. Our model has a per-atom representation module, a per-residue trunk that fuses the monomer features at its entrance, and a per-atom decoder that outputs the denosing predictions.

\textbf{Tokenization and monomer representation extraction.}
For each protein, we convert its structure into per-residue and per-atom token tables using the Boltz tokenizer~\citep{passaro2025boltz2,wohlwend2025boltz}, which standardizes residue indexing, atom typing, and bond information across the training pool. We then compute per-residue features for the monomer protein with the frozen representation model $h_\text{monomer}$. Across runs, we change only the monomer representation model. The atom-processing layers, trunk, and atom decoder are shared across all runs.

\textbf{Atom representation module and representation fusion.}
The atom representation module is a windowed Transformer with depth 1, hidden size 256, and 4 attention heads. This processes per-atom features, e.g., atom and bond types, together with per-atom coordinates of the current diffusion state $x_t$, and produces per-atom latents $a \in \mathbb{R}^{N_\text{atom} \times 256}$. We aggregate $a$ to per-residue latents via mean pooling along an atom-to-token assignment, and project up to the trunk hidden size of 768. 

Then, we construct the monomer input by concatenating two copies of $h$ along the residue dimension and adding learned chain-identity embeddings $E_\text{chain}$ that distinguish chain-A and chain-B positions:
\begin{equation}
\tilde{h} = \mathrm{concat}(h_\text{monomer}, h_\text{monomer}) + E_\text{chain} \in \mathbb{R}^{2L \times D_\text{monomer}}.
\end{equation}
We then fuse $\tilde{h}$ into the per-residue latent at the trunk entrance via concatenation followed by a learned linear layer $W_\text{cat}: \mathbb{R}^{768 + D_\text{monomer}} \to \mathbb{R}^{768}$:
\begin{equation}
z = W_\text{cat}\big(\mathrm{concat}(\text{token-latent}, \tilde{h})\big) \in \mathbb{R}^{2L \times 768}.
\end{equation}

\textbf{Trunk.} The trunk is an 8-layer DiT~\citep{peebles2023scalable} stack with hidden size 768, 12 attention heads (head dim 64), SwiGLU MLPs of ratio 4, and QK-norm along (chain, residue, atom, sub-atom) axes. The trunk operates only on the residue-level latents and outputs $z' \in \mathbb{R}^{2L \times 768}$.

\textbf{Per-atom decoder.}
We expand $z'$ back to per-atom latents through the atom-to-token assignment, residual-add the per-atom representation output $a$, project down to the decoder hidden size 256, and pass through a windowed Transformer decoder of depth 1, followed by a final AdaLN layer that produces the predicted velocity $v_\text{pred} \in \mathbb{R}^{2L \times N_\text{atom} \times 3}$.

\textbf{Flow-matching training.}
We train the model via flow matching. Given clean homodimer coordinates $x_1$ and noise $x_0 \sim \mathcal{N}(0, I)$, training algorithm samples $t \sim U(0, 1)$ and computes the interpolant $x_t = (1-t), x_0 + t, x_1$. Then, the model is trained to minimize:
\begin{equation}
\mathcal{L} = \mathbb{E}_{t,x_0,x_1} \big| v_{\theta}(x_t, t, \tilde{h}) - (x_1 - x_0) \big|^2 + \lambda_{\text{lDDT}}\, \mathcal{L}_\text{LDDT}^\text{smooth},
\end{equation}
where $\mathcal{L}_\text{LDDT}^\text{smooth}$ is a smooth lDDT auxiliary on the denoised coordinates. At inference, we sample $x_0 \sim \mathcal{N}(0, I)$ and integrate $v_\theta$ from $t = 0$ to $t = 1$ with an Euler-Maruyama integrator to obtain the predicted homodimer coordinates.

\textbf{Hyperparameters and experimental settings.} For all experiments, we use four NVIDIA B200 GPUs. We train the model for $100{,}000$ steps with AdamW using a learning rate of $2 \times 10^{-4}$ and weight decay of $0.01$. We use gradient clipping with norm $2.0$ and use norm $0.5$ for 2.8B representations to prevent exploding gradients. We use \texttt{bf16-mixed} precision. Following Boltz~\citep{wohlwend2025boltz,passaro2025boltz2}, we apply a multiplicity factor of $8$ to each sample, which replicates each sample with $8$ independent random SE(3) augmentations within a step. With $4$ samples per GPU and DDP across $4$ GPUs, the effective batch size is $128$ samples per step. During training, we crop each protein to $512$ residue tokens using the Boltz interface-aware cropper and pad to a static shape for \texttt{torch.compile}. The auxiliary smooth-lDDT loss weight is $\lambda_\text{lDDT} = 1.0$. At inference, we use $500$ time steps and noise scale $\tau = 0.3$ to obtain the predicted homodimer coordinates.

\subsection{Per-residue homodimer binding property prediction}\label[appendix]{appx:benchmark_perresidue}

\textbf{Label construction.} We first compute binding-relevant properties from the homodimer structures. These properties cannot be directly obtained from the monomer alone because they require the corresponding homodimer structure. To be specific, we derive four targets for residue $i$ as follows.

\emph{Binding-site prediction (binary classification).} A residue is labeled positive if its $\mathrm{C}_\alpha$ atom lies within $8$\,\AA{} of any $\mathrm{C}_\alpha$ atom on the partner chain.

\emph{$\Delta$SASA prediction (regression).} For the monomer and the homodimer, we compute the per-residue solvent-accessible surface area using the Shrake-Rupley algorithm~\citep{shrake1973environment} as implemented in \texttt{BioPython} \citep{cock2009biopython}, with probe radius $1.4$\,\AA{} and $960$ sphere points and get labels:
\[
\Delta\mathrm{SASA}_i \;=\; \mathrm{SASA}^{\text{mono}}_i - \mathrm{SASA}^{\text{dimer}}_i.
\]

\emph{Levy tier (multi-class classification).}
We assign each residue to one of five structural regions following the support-core-rim definition~\citep{levy2010simple}. This computes the relative SASA
$r_i = \mathrm{SASA}_i / \mathrm{SASA}^{\mathrm{max}}_{aa(i)}$, where $\mathrm{SASA}^{\mathrm{max}}_{aa(i)}$ is the maximum SASA for residue type $aa(i)$. Then, this computes $r_i$ for both the monomer and the homodimer. Residues without inter-chain contacts are assigned to \emph{surface} if $r_i^{\mathrm{mono}} > 0.25$ and to \emph{interior} if $r_i^{\mathrm{mono}} \leq 0.25$. Contacting residues are assigned to \emph{support} if they are already buried in the monomer ($r_i^{\mathrm{mono}} \leq 0.25$), to \emph{rim} if they remain exposed in both states ($r_i^{\mathrm{mono}} > 0.25$ and $r_i^{\mathrm{dimer}} > 0.25$), and to \emph{core} if they are exposed in the monomer but become buried in the dimer ($r_i^{\mathrm{mono}} > 0.25$ and $r_i^{\mathrm{dimer}} \leq 0.25$).

\emph{Bond type (multi-label classification).} For each inter-chain residue pair, we test five non-covalent interaction types: hydrogen bonds, salt bridges, hydrophobic contacts, $\pi$--$\pi$ stacking, and cation--$\pi$ interactions. We compute interaction labels using PLIP~\citep{salentin2015plip}. Residue-level labels are obtained by taking the union of interaction types over all inter-chain partners of each residue.

\textbf{Aggregation across chain instances.} 
Each residue position appears twice in the homodimer, once in each chain. Since the representation model takes the monomer as input and is chain-blind, we merge the two chain-specific labels into a single per-position label: OR for binding site and bond type, mean for $\Delta$SASA, and max-rank aggregation for Levy tier.

\textbf{Hyperparameters and experimental settings.}
We use a deterministic $10\%$ subset of the training split for probe training to better evaluate generalization capability of representations, while using the full validation ($400$) and test ($1{,}000$) sets. For each monomer protein, we extract residue-wise monomer representations and train a two-hidden-layer MLP with hidden width $1280$, GELU activations, and dropout $0.1$. We train each probe for $10$ epochs with a residue-wise batch size of $16{,}824$, which corresponds to approximately $40$ to $50$ proteins per batch.

\subsection{Monomer structure prediction through distillation}\label[appendix]{appx:benchmark_monomer}

\textbf{Architecture.}
Our folding model follows the atom-token-atom design of SimpleFold~\citep{wang_simplefold_2025}. A per-atom encoder maps atomic features together with the current diffusion state $x_t$ into atom latents, which are pooled into per-residue tokens. The tokens pass through a residue-level Transformer trunk, and a per-atom decoder reads the trunk output to predict the flow-matching velocity.

The trunk also incorporates per-residue embeddings from a frozen ESM2-650M sequence backbone~\citep{lin2023evolutionary}. These embeddings are added to the residue tokens at the trunk entrance. The trunk is an $8$-layer DiT~\citep{peebles2023scalable} stack with hidden size $768$, $12$ attention heads, head dimension $64$, SwiGLU MLPs with ratio $4$, and QK-norm. The atom encoder and decoder are windowed Transformers with depth $1$, hidden size $256$, and $4$ attention heads, identical to SimpleFold.


\textbf{Representation alignment.}
We evaluate whether frozen monomer representations provide useful auxiliary supervision for monomer structure prediction. For each representation model, we first extract residue-wise monomer protein representations as representation alignment targets $h_{\text{tgt}}$ and cache them. During training, we align the intermediate representation of the structure prediction model with the frozen monomer representation. Specifically, we use the output of the $4$-th DiT block's representation $z$, and align it to the cached target representation $h_{\text{tgt}}$. A two-layer MLP maps the source representation to the target representation dimension:
\begin{equation}
\hat{h}
=
g(z)
\end{equation}
where the hidden size of $g$ is $\max(768, D_{h_{\text{tgt}}})$. We then minimize a residue-wise negative cosine similarity between the projected source representation and the cached target representation:
\begin{equation}
\mathcal{L}_{\mathrm{REPA}}
=
-\lambda_{\mathrm{REPA}}\frac{1}{|\mathcal{V}|}
\sum_{i \in \mathcal{V}}
\frac{
\langle \hat{h}_i, h_{\mathrm{tgt}, i} \rangle
}{
\|\hat{h}_i\|_2 \, \|h_{\mathrm{tgt}, i}\|_2
},
\end{equation}
where $\mathcal{V}$ denotes the valid residue indices in the cropped sample. Gradients from $\mathcal{L}_{\mathrm{REPA}}$ are applied only to the DiT trunk and the projection head $g$.

\textbf{Hyperparameters and experimental settings.} We train on four NVIDIA B200 GPUs for $30{,}000$ steps with AdamW (learning rate $4 \times 10^{-4}$, weight decay $0.01$), gradient-norm clipping at $2.0$, and \texttt{bf16-mixed} precision. We crop with the Boltz cropper to $\mathrm{max\_tokens}=256$, padded to static shapes for \texttt{torch.compile}. We use a multiplicity factor of $8$, a per-GPU batch of $4$, and DDP across $4$ GPUs, giving an effective batch size of $128$ samples per step. The smooth-lDDT weight is $\lambda_{\mathrm{lDDT}} = 1.0$ and the REPA weight is $\lambda_{\mathrm{REPA}} = 2.0$. At inference, we use $500$ time steps and noise scale $\tau = 0.3$ to obtain the predicted monomer coordinates.

\newpage

\section{Additional experiments}\label[appendix]{appx:additional_exp}

\subsection{Additional per-residue property prediction}\label[appendix]{appx:full_perresidue}

\begin{table}[h]
\centering
\caption{\textbf{Per-residue homodimer binding properties probing on \bench.} MLP probes on monomer protein representations predict properties derived from the holodimer. \textbf{Bold} indicates the best within each size class. We use three random seeds.}
\label{tab:homomer-probing-std}
\centering
\resizebox{1.0\linewidth}{!}{
\setlength{\tabcolsep}{3pt}
\renewcommand{\arraystretch}{0.95}
\begin{tabular}{lc|cc|cc|cc|cc}
\toprule
& & \multicolumn{2}{c|}{\textbf{Binding site}}
& \multicolumn{2}{c|}{\textbf{$\Delta$SASA}}
& \multicolumn{2}{c|}{\textbf{Levy tier}}
& \multicolumn{2}{c}{\textbf{Bond type}} \\
\textbf{Models} & \textbf{Params}
& AUPRC & AUROC 
& Pearson & Spearman 
& Macro F1 & Acc. 
& AUPRC & AUROC \\
\hline
\rowcolor{gray!12}\multicolumn{10}{l}{\textit{Small}} \\
ESM2    & 35M    & 0.6635${\scriptstyle\pm 0.0009}$ & 0.8862${\scriptstyle\pm 0.0005}$ & 0.5955${\scriptstyle\pm 0.0015}$ & 0.3445${\scriptstyle\pm 0.0015}$ & 0.5386${\scriptstyle\pm 0.0027}$ & 0.7706${\scriptstyle\pm 0.0002}$ & 0.2936${\scriptstyle\pm 0.0004}$ & 0.9338${\scriptstyle\pm 0.0003}$ \\
SaProt  & 35M    & 0.6291${\scriptstyle\pm 0.0009}$ & 0.8794${\scriptstyle\pm 0.0005}$ & 0.6218${\scriptstyle\pm 0.0007}$ & 0.3923${\scriptstyle\pm 0.0010}$ & 0.5248${\scriptstyle\pm 0.0041}$ & 0.8036${\scriptstyle\pm 0.0003}$ & 0.2663${\scriptstyle\pm 0.0014}$ & 0.9388${\scriptstyle\pm 0.0001}$ \\
Ours    & 35M    & \textbf{0.7614}${\scriptstyle\pm 0.0003}$ & \textbf{0.9243}${\scriptstyle\pm 0.0002}$ & \textbf{0.7047}${\scriptstyle\pm 0.0003}$ & \textbf{0.4167}${\scriptstyle\pm 0.0006}$ & \textbf{0.6260}${\scriptstyle\pm 0.0037}$ & \textbf{0.8342}${\scriptstyle\pm 0.0003}$ & \textbf{0.3566}${\scriptstyle\pm 0.0002}$ & \textbf{0.9518}${\scriptstyle\pm 0.0002}$ \\
\hline
\rowcolor{gray!12}\multicolumn{10}{l}{\textit{Medium}} \\
ESM2    & 150M   & 0.7087${\scriptstyle\pm 0.0025}$ & 0.9062${\scriptstyle\pm 0.0009}$ & 0.6379${\scriptstyle\pm 0.0006}$ & 0.3703${\scriptstyle\pm 0.0014}$ & 0.5665${\scriptstyle\pm 0.0031}$ & 0.7969${\scriptstyle\pm 0.0003}$ & 0.3227${\scriptstyle\pm 0.0034}$ & 0.9425${\scriptstyle\pm 0.0003}$ \\
Ours    & 150M   & \textbf{0.8161}${\scriptstyle\pm 0.0004}$ & \textbf{0.9433}${\scriptstyle\pm 0.0002}$ & \textbf{0.7801}${\scriptstyle\pm 0.0003}$ & \textbf{0.4671}${\scriptstyle\pm 0.0012}$ & \textbf{0.6843}${\scriptstyle\pm 0.0021}$ & \textbf{0.8668}${\scriptstyle\pm 0.0001}$ & \textbf{0.4204}${\scriptstyle\pm 0.0024}$ & \textbf{0.9642}${\scriptstyle\pm 0.0002}$ \\
\hline
\rowcolor{gray!12}\multicolumn{10}{l}{\textit{Large}} \\
ESM2    & 650M   & 0.7915${\scriptstyle\pm 0.0006}$ & 0.9369${\scriptstyle\pm 0.0006}$ & 0.7146${\scriptstyle\pm 0.0007}$ & 0.4113${\scriptstyle\pm 0.0007}$ & 0.6442${\scriptstyle\pm 0.0033}$ & 0.8281${\scriptstyle\pm 0.0010}$ & 0.3902${\scriptstyle\pm 0.0020}$ & 0.9566${\scriptstyle\pm 0.0001}$ \\
SaProt  & 650M   & 0.8273${\scriptstyle\pm 0.0008}$ & 0.9508${\scriptstyle\pm 0.0003}$ & 0.7881${\scriptstyle\pm 0.0002}$ & 0.4715${\scriptstyle\pm 0.0012}$ & 0.6946${\scriptstyle\pm 0.0014}$ & 0.8693${\scriptstyle\pm 0.0004}$ & 0.4305${\scriptstyle\pm 0.0016}$ & \textbf{0.9692}${\scriptstyle\pm 0.0002}$ \\
S-PLM   & 704M   & 0.7789${\scriptstyle\pm 0.0017}$ & 0.9324${\scriptstyle\pm 0.0006}$ & 0.6967${\scriptstyle\pm 0.0002}$ & 0.4036${\scriptstyle\pm 0.0013}$ & 0.6215${\scriptstyle\pm 0.0021}$ & 0.8166${\scriptstyle\pm 0.0001}$ & 0.3602${\scriptstyle\pm 0.0024}$ & 0.9523${\scriptstyle\pm 0.0001}$ \\
MIF-ST  & 643M   & 0.5846${\scriptstyle\pm 0.0016}$ & 0.8671${\scriptstyle\pm 0.0007}$ & 0.6726${\scriptstyle\pm 0.0008}$ & 0.4338${\scriptstyle\pm 0.0011}$ & 0.5261${\scriptstyle\pm 0.0022}$ & 0.8435${\scriptstyle\pm 0.0002}$ & 0.2623${\scriptstyle\pm 0.0023}$ & 0.9430${\scriptstyle\pm 0.0003}$ \\
Ours    & 650M   & \textbf{0.8524}${\scriptstyle\pm 0.0011}$ & \textbf{0.9561}${\scriptstyle\pm 0.0002}$ & \textbf{0.8088}${\scriptstyle\pm 0.0001}$ & \textbf{0.4813}${\scriptstyle\pm 0.0005}$ & \textbf{0.7290}${\scriptstyle\pm 0.0009}$ & \textbf{0.8780}${\scriptstyle\pm 0.0001}$ & \textbf{0.4425}${\scriptstyle\pm 0.0050}$ & 0.9673${\scriptstyle\pm 0.0008}$ \\
\hline
\rowcolor{gray!12}\multicolumn{10}{l}{\textit{Huge}} \\
ESM2    & 3B     & 0.8247${\scriptstyle\pm 0.0008}$ & 0.9483${\scriptstyle\pm 0.0002}$ & 0.7446${\scriptstyle\pm 0.0009}$ & 0.4281${\scriptstyle\pm 0.0012}$ & 0.6851${\scriptstyle\pm 0.0054}$ & 0.8426${\scriptstyle\pm 0.0004}$ & 0.4106${\scriptstyle\pm 0.0027}$ & 0.9614${\scriptstyle\pm 0.0003}$ \\
ESM3    & 1.4B   & 0.8470${\scriptstyle\pm 0.0016}$ & 0.9558${\scriptstyle\pm 0.0005}$ & 0.8121${\scriptstyle\pm 0.0009}$ & \textbf{0.4878}${\scriptstyle\pm 0.0010}$ & 0.7285${\scriptstyle\pm 0.0048}$ & \textbf{0.8898}${\scriptstyle\pm 0.0002}$ & 0.4229${\scriptstyle\pm 0.0056}$ & 0.9673${\scriptstyle\pm 0.0002}$ \\
ProstT5 & 1.2B   & 0.8291${\scriptstyle\pm 0.0013}$ & 0.9480${\scriptstyle\pm 0.0004}$ & 0.7574${\scriptstyle\pm 0.0011}$ & 0.4449${\scriptstyle\pm 0.0004}$ & 0.6848${\scriptstyle\pm 0.0025}$ & 0.8538${\scriptstyle\pm 0.0005}$ & 0.4077${\scriptstyle\pm 0.0040}$ & 0.9594${\scriptstyle\pm 0.0002}$ \\
Ours    & 2.8B   & \textbf{0.8650}${\scriptstyle\pm 0.0007}$ & \textbf{0.9608}${\scriptstyle\pm 0.0005}$ & \textbf{0.8152}${\scriptstyle\pm 0.0009}$ & 0.4854${\scriptstyle\pm 0.0020}$ & \textbf{0.7408}${\scriptstyle\pm 0.0016}$ & 0.8811${\scriptstyle\pm 0.0003}$ & \textbf{0.4516}${\scriptstyle\pm 0.0017}$ & \textbf{0.9691}${\scriptstyle\pm 0.0011}$ \\
\bottomrule
\end{tabular}}
\end{table}

\subsection{Additional results on monomer structure prediction}\label[appendix]{appx:huge_apo_repa}

\begin{table}[h]
\centering
\caption{\textbf{Additionnal results on monomer structure prediction with disillation on \bench{}.} \textbf{Bold} denotes the best performance.}\label{tab:huge_apo_repa}
\centering
\scalebox{0.9}{
\setlength{\tabcolsep}{4pt}
\renewcommand{\arraystretch}{1.15}
\begin{tabular}{lc|cccccc}
\toprule
\textbf{REPA target} & \textbf{Params}
& TM-score $\uparrow$ & GDT-TS $\uparrow$ & GDT-HA $\uparrow$
& LDDT $\uparrow$ & LDDT$_\mathrm{bb}$ $\uparrow$ & RMSD (\AA) $\downarrow$ \\
\midrule
Baseline & --   & 0.5422 & 0.3432 & 0.1990 & 0.6208 & 0.6746 & 12.98 \\
SaProt   & 650M & 0.5659 & 0.3677 & 0.2192 & 0.6410 & 0.6947 & 12.26 \\
S-PLM    & 704M & 0.5569 & 0.3559 & 0.2081 & 0.6301 & 0.6838 & 12.40 \\
MIF-ST   & 643M & 0.5491 & 0.3520 & 0.2075 & 0.6310 & 0.6827 & 12.73 \\
ESM3     & 1.4B & 0.5684 & 0.3714 & 0.2223 & 0.6366 & 0.6900 & 12.25 \\
ProstT5  & 1.2B & \textbf{0.5833} & 0.3854 & 0.2328 & \textbf{0.6535} & \textbf{0.7119} & \textbf{11.76} \\
Ours     & 650M & 0.5822 & \textbf{0.3858} & \textbf{0.2344} & 0.6513 & 0.7051 & 11.91 \\
\bottomrule
\end{tabular}}
\end{table}

\subsection{Per-residue heterodimer binding-property probing}
\label[appendix]{appx:benchmark_heterodimer_balanced}

\textbf{Dataset curation.} We probe per-target-residue binding properties on
heterodimers from PPIRef50K~\citep{pipref50k}, restricting to entries
where the binder and target chains have non-identical sequences
($n=15{,}913$ complexes; sequence-identical homodimers are excluded so
that labels strictly reflect cross-chain interactions between distinct
proteins). Each complex consists of two distinct chains (binder,
target), so labels are not aggregated across chain instances. The
representation model takes the target monomer as input, and we evaluate
two complementary target-conditioned labels: a binary
\emph{binding-site} indicator and a continuous \emph{contact count}.

\textbf{Binding-site prediction (binary classification).} Per-residue
solvent-accessible surface area is computed for both the target monomer
and the binder--target complex with DSSP~\citep{dssp}, using the
Sander--Rost relative-ASA normalization. Let $r^{\mathrm{mono}}_i$ and
$r^{\mathrm{cpx}}_i$ denote the relative ASA of target residue $i$ in
the monomer and the complex, respectively, and define the burial change
$\Delta r_i = r^{\mathrm{mono}}_i - r^{\mathrm{cpx}}_i$. A residue is
labeled positive if $\Delta r_i \geq 0.05$ and both $r^{\mathrm{mono}}_i$,
$r^{\mathrm{cpx}}_i$ are finite, i.e., its surface is measurably occluded
by the binder. Residues with non-finite DSSP output (missing atoms,
alternative locations) are dropped from training and evaluation.

\textbf{Contact-count prediction (regression).} For each target residue
$i$ we count the number of partner $\mathrm{C}_\alpha$ atoms within
$8$\,\AA{} of $i$'s $\mathrm{C}_\alpha$:
\begin{equation}
c_i =
\left|
\left\{
j \in \mathrm{binder}
:
\left\|
\mathrm{C}^{t}_{\alpha,i} - \mathrm{C}^{b}_{\alpha,j}
\right\| \le 8\,\text{\AA}
\right\}
\right|.
\label{eq:target_contact_count}
\end{equation}
Compared with the binary binding-site label, $c_i$ retains a graded
signal of how deeply each interface residue is packed against the
partner, capturing local contact density and hotspot strength. Targets
are stored as $\log(1+c_i)$ and the probe is trained with mean-squared
error on the transformed target; we report Pearson $r$ and Spearman
$\rho$ on the original count scale.

\textbf{Class-balanced subsampling.} Across heterodimer targets, only
${\sim}10.9\%$ of residues are interface residues, which biases probes
toward majority-class collapse for the binary task and zero-inflation
bias for the regression task (a trivial $\hat c_i = 0$ predictor would
attain a low MSE while learning nothing). To control prevalence we
construct a single deterministic class-balanced subsample that we apply
uniformly to both tasks: we keep \emph{all} interface residues
($\Delta r_i \geq 0.05$) and globally Bernoulli-sample non-interface
residues with probability $p = n_{\text{interface}} / n_{\text{non}}$,
yielding $\approx 50/50$ prevalence ($\sim 620{,}000$ residues kept).
The same per-PPI keep-mask is applied identically to the training,
validation, and test splits (the split is PDB-level, so a target's
residue indices retain their split tag while the mask determines
residue-level inclusion). The mask is generated once with a fixed seed
(default $42$) so that all PLMs see exactly the same residues across
the binding-site and contact-count tasks. Because train and test share
the same balancing rule, chance baselines shift accordingly: AUPRC
chance is $0.5$ for binding-site, and the variance of $\log(1+c_i)$ on
the balanced subset replaces the zero-inflated baseline for regression.

\textbf{Hyperparameters and experimental settings.} For each target
monomer, we extract residue-wise representations from the frozen PLM
and feed them into a probe head. The probe head is a two-hidden-layer
MLP with hidden width $1{,}280$, GELU activations, and dropout $0.1$,
optimized with AdamW (weight decay $0.01$). We train each probe for
$10$ epochs with a residue-wise batch size of $16{,}824$ and a learning
rate of $5\!\times\!10^{-4}$. Binding-site uses cross-entropy with
uniform class weights (the subsample already removes majority-class
dominance, so reweighting would collapse to identity); contact-count
uses mean-squared error on $\log(1+c_i)$. We evaluate on the full
validation and test splits (residues filtered by the same balanced
mask). Binding-site is reported with AUPRC and AUROC; contact-count
with Pearson $r$ and Spearman $\rho$.

\begin{table}[h]
\centering
\caption{\textbf{Per-residue heterodimer binding properties probing on \bench.} MLP probes on monomer protein representations predict properties derived from the heterodimer complex. Both labels are target-conditioned and evaluated on a class-balanced subset. \textbf{Bold} indicates the best performance within each model-size group. Our model achieves the strongest performance.}
\label{tab:binder-screening-binding-site-full}
\centering
\scalebox{0.9}{
\setlength{\tabcolsep}{4pt}
\renewcommand{\arraystretch}{1.15}
\begin{tabular}{lc|cc|cc}
\toprule
& & \multicolumn{2}{c|}{\textbf{Binding site}}
& \multicolumn{2}{c}{\textbf{Contact count}} \\
\textbf{Models} & \textbf{Params}
& AUPRC & AUROC
& Pearson & Spearman \\
\hline
\rowcolor{gray!12}\multicolumn{6}{l}{\textit{Small}} \\
ESM-2        & 35M    & $0.717{\scriptstyle \pm 0.001}$ & $0.656{\scriptstyle \pm 0.002}$ & $0.339{\scriptstyle \pm 0.002}$ & $0.322{\scriptstyle \pm 0.001}$ \\
Ours         & 35M    & $\mathbf{0.755{\scriptstyle \pm 0.001}}$ & $\mathbf{0.693{\scriptstyle \pm 0.001}}$ & $\mathbf{0.411{\scriptstyle \pm 0.002}}$ & $\mathbf{0.401{\scriptstyle \pm 0.002}}$ \\
\hline
\rowcolor{gray!12}\multicolumn{6}{l}{\textit{Medium}} \\
ESM-2        & 150M   & $0.741{\scriptstyle \pm 0.001}$ & $0.676{\scriptstyle \pm 0.001}$ & $0.376{\scriptstyle \pm 0.003}$ & $0.362{\scriptstyle \pm 0.003}$ \\
Ours         & 150M   & $\mathbf{0.796{\scriptstyle \pm 0.001}}$ & $\mathbf{0.727{\scriptstyle \pm 0.002}}$ & $\mathbf{0.460{\scriptstyle \pm 0.004}}$ & $\mathbf{0.445{\scriptstyle \pm 0.004}}$ \\
\hline
\rowcolor{gray!12}\multicolumn{6}{l}{\textit{Large}} \\
ESM-2        & 650M   & $0.759{\scriptstyle \pm 0.002}$ & $0.692{\scriptstyle \pm 0.001}$ & $0.407{\scriptstyle \pm 0.002}$ & $0.391{\scriptstyle \pm 0.002}$ \\
SaProt       & 650M   & $0.797{\scriptstyle \pm 0.001}$ & $0.730{\scriptstyle \pm 0.001}$ & $\mathbf{0.500{\scriptstyle \pm 0.002}}$ & $\mathbf{0.476{\scriptstyle \pm 0.002}}$ \\
S-PLM        & 704M   & $0.755{\scriptstyle \pm 0.001}$ & $0.690{\scriptstyle \pm 0.001}$ & $0.402{\scriptstyle \pm 0.002}$ & $0.390{\scriptstyle \pm 0.001}$ \\
MIF-ST       & 643M   & $0.757{\scriptstyle \pm 0.000}$ & $0.695{\scriptstyle \pm 0.001}$ & $0.448{\scriptstyle \pm 0.002}$ & $0.435{\scriptstyle \pm 0.003}$ \\
Ours         & 650M   & $\mathbf{0.817{\scriptstyle \pm 0.001}}$ & $\mathbf{0.744{\scriptstyle \pm 0.000}}$ & $0.482{\scriptstyle \pm 0.001}$ & $0.468{\scriptstyle \pm 0.001}$ \\
\hline
\rowcolor{gray!12}\multicolumn{6}{l}{\textit{Huge}} \\
ESM-2        & 3B     & $0.764{\scriptstyle \pm 0.000}$ & $0.697{\scriptstyle \pm 0.000}$ & $0.415{\scriptstyle \pm 0.002}$ & $0.396{\scriptstyle \pm 0.002}$ \\
ESM3         & 1.4B   & $0.746{\scriptstyle \pm 0.007}$ & $0.696{\scriptstyle \pm 0.004}$ & $0.047{\scriptstyle \pm 0.007}$ & $0.045{\scriptstyle \pm 0.010}$ \\
ProSST       & 1.2B   & $0.761{\scriptstyle \pm 0.000}$ & $0.694{\scriptstyle \pm 0.000}$ & $0.442{\scriptstyle \pm 0.002}$ & $0.426{\scriptstyle \pm 0.002}$ \\
Ours         & 2.8B  & $\mathbf{0.834{\scriptstyle \pm 0.001}}$ & $\mathbf{0.761{\scriptstyle \pm 0.000}}$ & $\mathbf{0.517{\scriptstyle \pm 0.000}}$ & $\mathbf{0.502{\scriptstyle \pm 0.000}}$ \\
\bottomrule
\end{tabular}}
\end{table}

\subsection{Conventional benchmark}\label[appendix]{appx:conventional_benchmark}

Following SaProt~\citep{su2024saprot}, instead of native crystal structures we use AlphaFold predicted structures from the AFDB corresponding to each protein sequence. For each protein, we extract last-layer residue-wise representations from the pretrained encoder and mean-pool over non-special tokens to obtain a single per-protein embedding. We then train a two-hidden-layer MLP probe with hidden width $1280$, GELU activations, and dropout $0.1$, using binary cross-entropy. Each probe is trained for $100$ epochs with Adam (learning rate $5\times 10^{-4}$) and a per-protein batch size of $128$. We select the epoch with the highest validation $F_{\max}$ and report the corresponding test $F_{\max}$.









\end{document}